\documentclass[aps,prl,amsfonts,amssymb,twocolumn,amsmath,preprintnumbers,floatfix,superscriptaddress]{revtex4-2}%
\usepackage{appendix}
\usepackage{graphicx}
\usepackage{mathrsfs}
\usepackage{bm}
\usepackage{amsmath}
\usepackage{amssymb}
\usepackage{braket}
\usepackage{ulem}
\usepackage[colorlinks,linkcolor=blue,anchorcolor=blue,citecolor=blue,urlcolor=blue]{hyperref}
\hypersetup{hypertex=true,
            colorlinks=true,
            linkcolor=blue,
            anchorcolor=blue,
            citecolor=blue}

\begin{document}

\newcommand{\mb}{\mathbf}
\newcommand{\mc}{\mathcal}
\newcommand{\bs}{\boldsymbol}


\title{Topological Charge-Transfer Excitons}

\author{Huiyuan Zheng}
\email{hyzheng@uw.edu}
\affiliation{Department of Materials Science and Engineering, University of Washington, Seattle, Washington 98195, USA}

\author{Kaijie Yang}
\affiliation{Department of Materials Science and Engineering, University of Washington, Seattle, Washington 98195, USA}

\author{Ting Cao}
\email{tingcao@uw.edu}
\affiliation{Department of Materials Science and Engineering, University of Washington, Seattle, Washington 98195, USA}

\author{Di Xiao}
\email{dixiao@uw.edu}
\affiliation{Department of Materials Science and Engineering, University of Washington, Seattle, Washington 98195, USA}
\affiliation{Department of Physics, University of Washington, Seattle, Washington 98195, USA}



\begin{abstract}
Excitons possess internal structure absent from single-particle Bloch particles, allowing their band topology to emerge from the bound-state structure rather than being inherited from their constituents.
This raises the question of how the internal structure of a bound state can provide a microscopic origin of exciton topology.
Here we show that the real-space embedding of charge-transfer excitons can generate an intrinsic manifold of symmetry-related off-site composite orbitals whose coupling supports topological exciton bands.
Lateral electron-hole separation embeds the localized exciton on the bond connecting its constituent sites rather than on either site. 
We demonstrate this mechanism in a honeycomb lattice, where three bond-centered charge-transfer exciton orbitals form a Kagome lattice. 
By solving the Bethe-Salpeter equation, we show that this emergent multi-orbital manifold supports a topological exciton flat band upon time-reversal symmetry breaking, even when the electron and hole bands are topologically trivial. 
The resulting band exhibits nearly uniformly distributed quantum geometry, favorable for interaction-driven bosonic states.
Our results establish a general route toward topological bands of localized composite bound states and unconventional correlated bosonic phases.
\end{abstract}

\maketitle

\textit{Introduction}---
In crystals, electron topology has become a central framework for connecting lattice symmetry, orbital embedding, and wave-function geometry to macroscopic quantum phenomena~\cite{fu2011topological, fang2012bulk,bradlyn2017topological}.
Composite particles, instead, introduce an additional route to nontrivial topology, because their motion is described not only by a center-of-mass (COM) coordinate but also by the relative configuration of their constituents.
Excitons, bound states of an electron and a hole, provide a simple two-body example in which this distinction can become consequential.
For the most commonly studied Rydberg-like excitons, strong electron-hole overlap ties the center of the exciton orbital to the underlying electronic sites~\cite{haug2009quantum,chernikov2014exciton}. 
Existing routes toward nontrivial exciton topology have therefore largely built on constituent-band topology~\cite{lozano2025optical,froese2025topological,kwan2021exciton,yang2026giant}, valley~\cite{wu2017topological,zheng2025forster} and layer~\cite{xie2024long} degrees of freedom, or on momentum-space envelope functions, such as hybridized Rydberg states~\cite{zhang2025engineering} and interaction-structured exciton wave functions~\cite{davenport2024interaction,jankowski2025excitonic}.
This leaves open a distinct possibility: the real-space arrangement of the electron and hole may endow the bound state with a crystalline orbital manifold distinct from either constituent.

Charge-transfer (CT) excitons, in which the electron and hole reside at laterally separated sites, provide a natural realization of this possibility.
While CT excitons are familiar from donor-acceptor materials~\cite{rosati2023interface,shradha20262d,choi1964charge,pereira2019electroabsorption}, 
van der Waals moir\'e superlattices now provide a tunable crystalline setting in which such electron-hole separation is organized by moir\'e periodicity~\cite{naik2022intralayer,wang2025twist,li2024imaging,wang2023intercell,zeng2022strong}.
In this setting, the electron-hole separation displaces the exciton COM away from either constituent orbital and allows multiple off-site CT configurations.
These off-site CT orbitals are related by crystal symmetry and become energetically comparable, thereby forming an intrinsic structure absent in the Rydberg regime.
The resulting multi-orbital manifold, emerging from the CT embedding of the composite bound state, provides a microscopic route to exciton topology that requires neither nontrivial topology in the constituent bands~\cite{lozano2025optical,froese2025topological,kwan2021exciton} nor a specially structured electron-hole interaction~\cite{davenport2024interaction}.

Building on this observation, we show how laterally
separated electrons and holes can be used to construct a topological exciton manifold.
The inequivalent CT orbitals act as an internal degree of freedom and give rise to exciton Berry curvature. 
We first illustrate the real-space embedding constraint and then show that a honeycomb arrangement, 
with the electron and hole residing on opposite sublattices, yields a Kagome configuration of CT excitons. 
By solving the Bethe-Salpeter equation, we obtain a topological exciton flat band from topologically trivial electron and hole bands, stabilized close to a combined particle-hole (PH) and inversion symmetry.
Our work identifies a generic route to engineering bosonic systems with nontrivial topology and quantum geometry from composite electron-hole bound states, thereby offering a promising platform for the realization of unconventional correlated bosonic phases.

\textit{Symmetry-enforced orbitals in CT excitons}---
We begin by examining the real-space structure of a localized exciton. 
Consider a composite orbital formed by an electron-hole pair trapped within a single unit cell.
For Rydberg-like excitons, the electron and hole largely overlap, so the exciton orbital is naturally centered on an underlying electronic site, and its topology is typically inherited from the constituent bands. 
CT excitons provide a fundamentally different route: when the electron and hole occupy distinct intracell positions, their separation creates a localized composite orbital whose center is shifted away from either constituent site. 
This relocation is not merely a geometric shift: it generically produces multiple CT configurations related by crystal symmetry, thereby forming a multi-orbital manifold.

\begin{figure}
\includegraphics[width=0.48\textwidth]{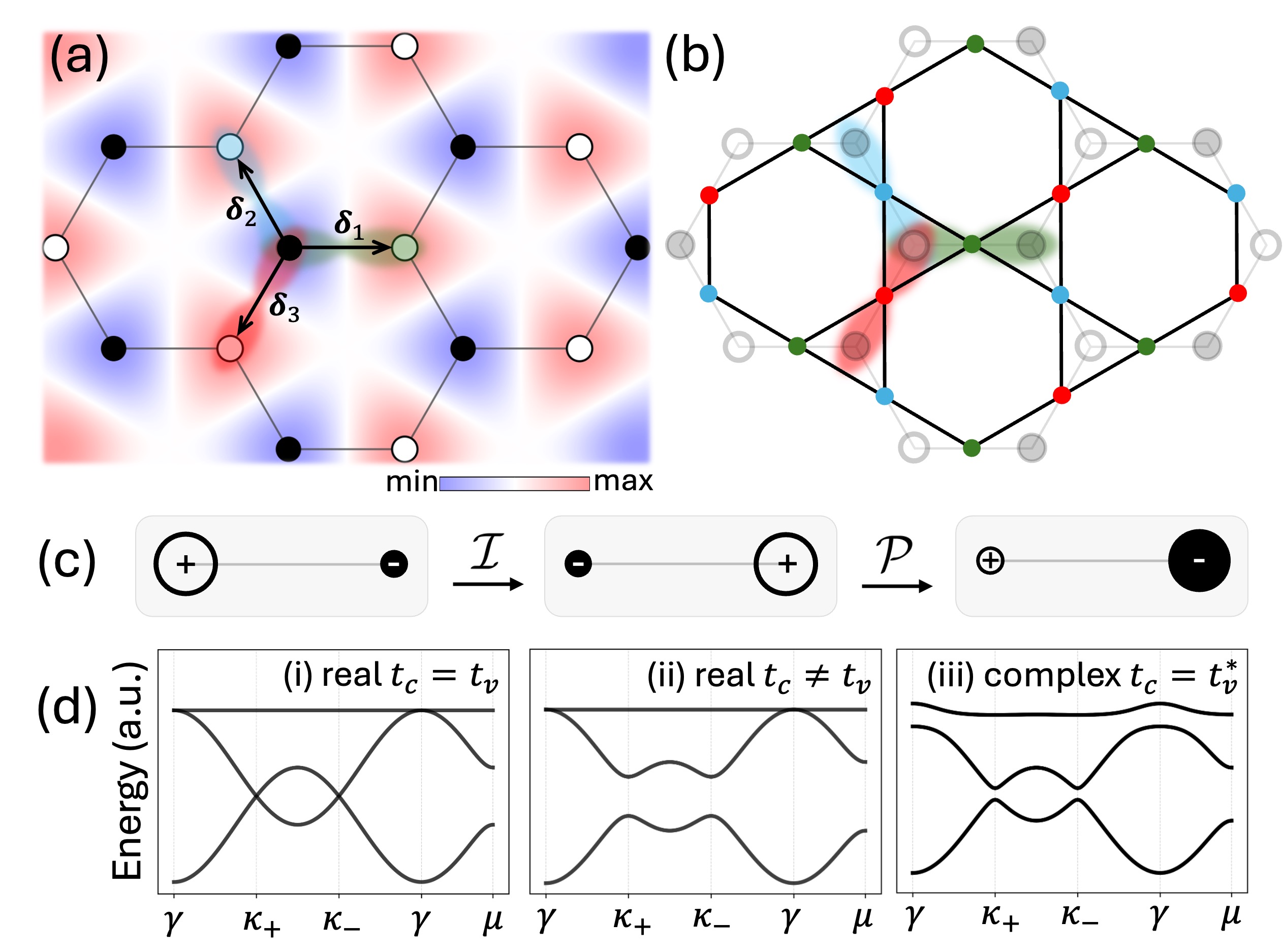}%
\caption{
(a) Schematic electron-hole distribution, with the electron (hole) denoted by a black (hollow) dot. $\bs{\delta}_{j=1,2,3}$ are the three relative coordinates of the electron-hole pair. 
The background color represents the trapping potential $V(\mb r)$.
(b) Schematic Kagome lattice of CT excitons. Green, blue, and red denote the three inequivalent orbitals.
The marker at each bond center is a schematic label for the symmetry center of the corresponding electron–hole bond configuration.
(c) Schematic CT exciton under a $\mc{PI}$ transformation. The electron (hole) is denoted by a solid (hollow) circle with negative (positive) charge. We use different circle sizes to illustrate asymmetric electronic orbitals.
(d) Exciton dispersion of Hamiltonian~\eqref{eq:KagomeTB}, calculated for (i) real $t_c=t_v$; (ii) real $t_c\neq t_v$; (iii) complex $t_c = t_v^*$.
\label{fig_Schematic}}
\end{figure}

To demonstrate this picture explicitly, we consider a honeycomb configuration in which the electron and hole occupy separate sublattices [Fig.~\ref{fig_Schematic}(a)].
In the strong-modulation limit, the trapping potentials become infinitely deep, so the electron and hole are well described by maximally localized Wannier orbitals $w_{c/v} (\mb r_{e/h})$, where $c$ and $v$ label the lowest conduction and highest valence Bloch bands, respectively.
In this limit, the excitons are described by the Wannier equation, and their eigenstates are simple products of electron and hole Wannier orbitals (we neglect the exchange interaction for simplicity),
$X_{j,\mb{r}_i}(\mb{r}_e, \mb{r}_h)=w_c(\mb{r}_e-\mb{r}_i-\bs{\delta}_j)w_v^*(\mb{r}_h-\mb{r}_i)$, where $\mb{r}_i$ is the common unit-cell coordinate of the CT exciton and $\bs{\delta}_{j=1,2,3}$ denote the three inequivalent relative coordinates [see details in the Supplemental Materials (SM)~\cite{supp}\nocite{zhao2021universal,kim2025moire,chiu2016classification}]. 
Turning on the electron and hole hopping perturbatively and retaining only nearest-neighbor hopping, yields an effective tight-binding Hamiltonian for CT excitons,
\begin{eqnarray}
    H_{\text{eff}} & = & \sum_i \sum_{j=1}^3 t_c  X^\dagger_{j+1,\mb{r}_i}X_{j,\mb{r}_i} \nonumber \\
    && \quad\quad \,\,\, + t_v X^\dagger_{j,\mb{r}_i}X_{j+1,\mb{r}_i-\mb{d}_{j+2}} + h.c.,
    \label{eq:KagomeTB}
\end{eqnarray}
where $X^\dagger_{j,\mb{r}_i}$ creates the exciton state $\ket{X_{j,\mb{r}_i}}=X^\dagger_{j,\mb{r}_i}\ket{\mathrm{GS}}$, and $\ket{\mathrm{GS}}$ denotes the ground state. 
We impose the cyclic condition $j+3 \equiv j\,\, (\mathrm{mod}\,\, 3)$. The three lattice vectors are $\mb{d}_{j+2}=\bs{\delta}_{j+1} - \bs{\delta}_{j}$.
Microscopically, each CT exciton orbital is formed by an electron and a hole on opposite sublattices, producing a finite electric dipole characterized by one of the three relative coordinates $\bs{\delta}_{j}$. 
The conduction-electron hopping $t_c$ converts an exciton orbital $\bs \delta_j$ into $\bs \delta_{j+1}$ within the same unit cell while leaving the valence-electron (hole) fixed. 
Similarly, the valence-electron hopping $t_v$ converts the exciton orbital $\bs \delta_{j+1}$ in the adjacent unit cell displaced by $-\mb d_{j+2}$ into $\bs \delta_j$ while leaving the conduction-electron fixed.
These three exciton orbitals are degenerate under the threefold rotation $\mc{C}_{3z}$.

We begin by assuming real and equal hopping amplitudes $t_c=t_v$.
Hamiltonian~\eqref{eq:KagomeTB} then reduces to a Kagome model of CT excitons, with two Dirac cones at the Brillouin zone (BZ) corners and one flat band [Fig.~\ref{fig_Schematic}d(i)].
A real-space interpretation provides an intuitive understanding of this Kagome nature. 
For $t_c=t_v$, $H_{\rm eff}$ is invariant under combined PH and inversion ($\mc{PI}$) symmetry.
For a localized CT exciton $X^\dagger_{j,\mb{r}_i}\ket{\mathrm{GS}}$, 
spatial inversion $\mc I$ exchanges two sublattice positions thus reversing the relative displacement from $\bs \delta_j$ to $-\bs \delta_j$ [Fig.~\ref{fig_Schematic}(c)].
By contrast, PH operation $\mc P$ is charge-conjugating and exchanges the conduction- and valence-orbital characters at fixed spatial coordinates.
Therefore, the combined $\mc{PI}$ operation maps the exciton operator
$(\mc{PI}) X^\dagger_{j,\mb{r}_i} (\mc{PI})^{-1}=X^\dagger_{j,-\mb{r}_i-\bs\delta_j}$ [see details in the SM~\cite{supp}].
In the original state, the hole occupies the site $\mb r_i$ and the electron occupies the neighboring site $\mb r_i+\bs\delta_j$. 
After the $\mc{PI}$ transformation, they occupy the inverted pair of sites, $-\mb r_i-\bs\delta_j$ and $-\mb r_i$, with their conduction- and valence-orbital roles exchanged. 
Thus, $\mc{PI}$ preserves the relative coordinate ($\bs \delta_j$), although it reverses the spatial arrangement of the two constituents [Fig.~\ref{fig_Schematic}(c)].
The bond center is thus a symmetry center of the localized pair configuration under $\mc{PI}$ symmetry.
The resulting low-energy degrees of freedom are the three symmetry-related CT exciton configurations in each unit cell, whose lattice connectivity is Kagome [Fig.~\ref{fig_Schematic}(b)].
This CT exciton lattice can be understood from the perspective of line graph~\cite{tasaki2020physics}. The constituent electron and hole serve as vertices, the CT configurations form bonds, two bonds are connected if they share a same vertex. If the nearest-neighbor hopping is restricted by $\mc{PI}$ symmetry, a flat band must exist.

For real $t_c=t_v$, the Kagome spectrum contains two Dirac points between the first and second bands and a quadratic touching between the second and third bands. 
These band-touching points are protected by $\mc{PI}$ and time reversal ($\mc T$) symmetries [Fig.~\ref{fig_Schematic}d(i)].
By allowing complex and unequal $t_c$ and $t_v$, band gaps can be opened. 
At the quadratic touching point $\gamma$, $\Delta_{\gamma} = |2\sqrt{3}\,\mathrm{Im}(t_c-t_v)|$,
the gap is opened only by $\mc{T}$ breaking, whereas $\mc{PI}$ breaking has no effect.
At Dirac points $\kappa_{\tau=\pm}$, $\Delta_{\kappa_{\tau}} = |3\mathrm{Re}(t_c-t_v) - \tau \sqrt{3} \mathrm{Im}(t_c-t_v)|$, 
so the gaps can be opened by breaking either $\mc T$ or $\mc{PI}$.
When $\mc{PI}$ symmetry is broken, a topologically trivial gap opens at the Dirac points [Fig.~\ref{fig_Schematic}d(ii)].
This result can be understood by adiabatically shifting the exciton toward the hole site, where the system becomes topologically equivalent to a triangular lattice of Rydberg-like excitons.
By contrast, the gap opened by $\mc{T}$ breaking within this Kagome configuration (with preserved $\mc{PI}$) is topologically nontrivial [Fig.~\ref{fig_Schematic}d(iii)].
Therefore, the topological gap of the exciton ground state is dictated by the competition between $\mc{T}$ breaking and $\mc{PI}$ breaking.

\textit{Exciton dispersion from BSE}---We now demonstrate the above idea in a concrete model.
The nontrivial exciton topology discussed above requires broken time reversal symmetry, which arises naturally from the massive Dirac cone $H_{\text{Dirac}}=\hbar v_F \mb{k} \cdot \bs{\sigma} +(\Delta_g/2)\sigma_z$ in valley-polarized moir\'e superlattices~\cite{xiao2012coupled,li2024imaging, wang2023intercell,naik2022intralayer,maity2026origin,wang2025twist,zhao2021universal,kim2025moire}. 
Here $\hbar$ is the reduced Planck constant, $v_F$ is the Fermi velocity, $\Delta_g$ is a large intrinsic gap separating the conduction ($c$) and valence ($v$) bands, and
$\bs{\sigma}$ is the vector of Pauli matrices operating in sublattice space.
Meanwhile, a spatially varying potential $V(\mb r)$ can stabilize the localized CT configuration. 
Motivated by these considerations, we introduce a continuum model that realizes the physical picture developed above:
\begin{equation}
    H_0= H_{\text{Dirac}}+ V(\mb{r}) \;.
    \label{eq:sp_hamiltonian}
\end{equation}

\begin{figure}
\includegraphics[width=0.48\textwidth]{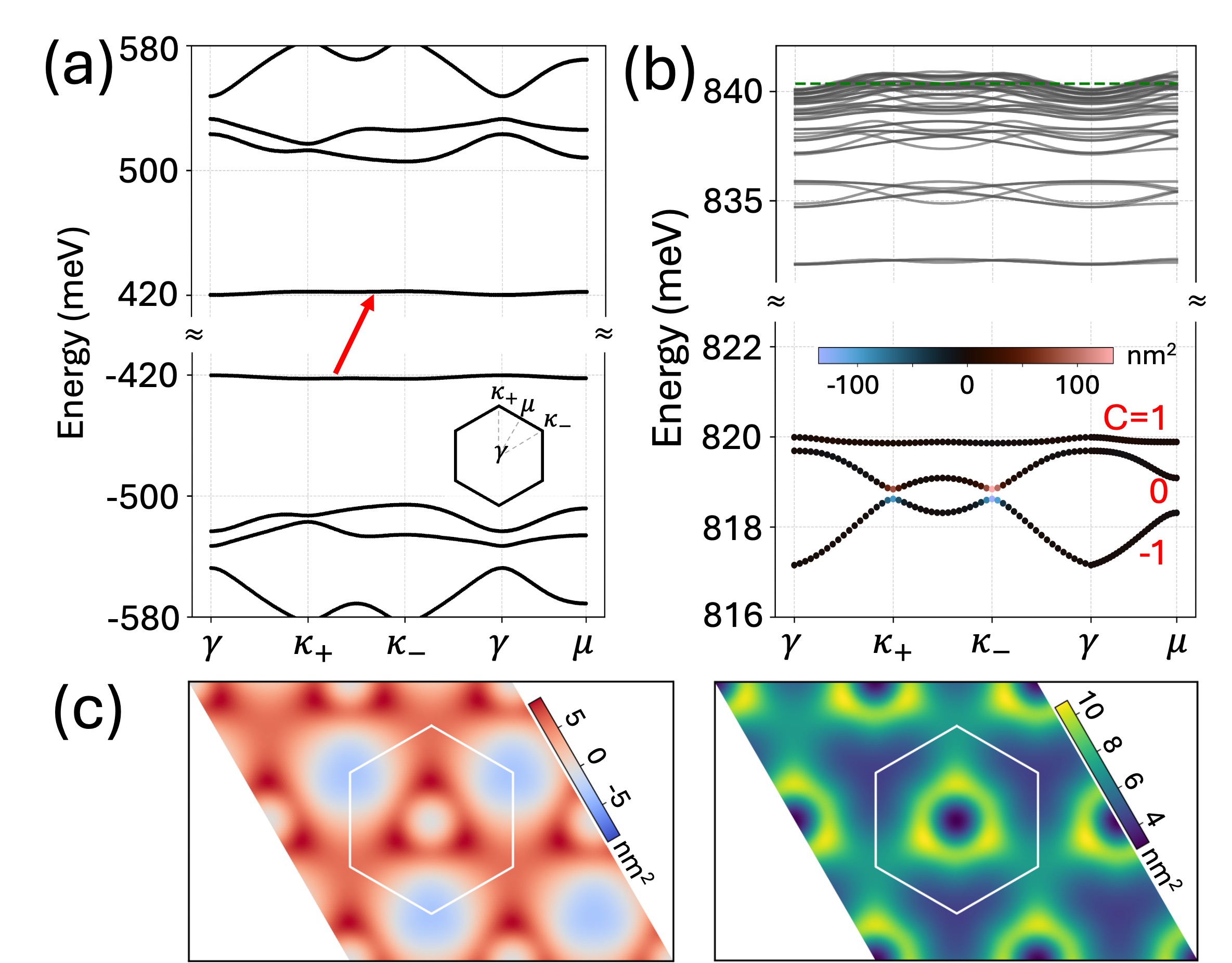}%
\caption{
(a) Single-particle dispersion from Hamiltonian~\eqref{eq:sp_hamiltonian}. The red arrow denotes excitation from the valence band to the conduction band. 
The calculation uses a 5 nm lattice constant, an intrinsic gap $\Delta_g=1$eV, and a Fermi velocity $v_F=3.3\times 10^5$m/s. 
(b) Exciton dispersion showing the lowest 50 BSE bands. 
The colors of the three lowest bands encode the exciton Berry curvature, with their Chern numbers marked in red. 
The green dashed line marks the single-particle gap.  
(c) Momentum-space distributions of the exciton Berry curvature (left) and quantum metric trace (right) for the exciton flat (third lowest) band in (b). The white honeycomb outlines the BZ.
\label{fig_Bse}}
\end{figure}

We expand the potential in the first harmonic shell
$V(\mb{r})=2\Delta\sum_{j=1,2,3}\cos(\mb{g}_j \cdot \mb{r} - \varphi)$, where $\Delta=40$meV is the potential strength and $\varphi$ is its phase; $\mb{g}_{1,2}$ are the primitive reciprocal-lattice vectors and $\mb{g}_{3}=-(\mb{g}_1 + \mb{g}_2)$. 
The odd potential obtained for $\varphi=\pi/2$ guarantees that Hamiltonian~\eqref{eq:sp_hamiltonian} is invariant under $\mc{PI}$ symmetry.
This potential landscape generates a pair of topologically trivial conduction and valence bands well isolated from the remote bands [Fig.~\ref{fig_Bse}(a)].
The lowest $c$-band states are localized on a triangular lattice, with their orbitals trapped at the potential minima of $V(\mb{r})$. 
Similarly, the highest $v$-band states have orbitals trapped at the potential maxima of $V(\mb{r})$ [Fig.~\ref{fig_Schematic}(a)]. 
The electron and hole therefore occupy opposite sublattices and together form a honeycomb lattice.

For the laterally separated $c$- and $v$-band orbitals, we obtain the exciton spectrum by solving the Bethe-Salpeter equation (BSE) throughout the BZ (see details in the SM~\cite{supp}). 
Because the lowest $c$-band and the highest $v$-band are well isolated, we retain only this pair and omit the band indices in the exciton basis.
Using the exciton basis $b^\dagger_{\mb Q,\mb k}=e^\dagger_{c,\mb k+\mb Q}e_{v,\mb k}$, with COM momentum $\mb Q$ and relative momentum $\mb k$, we write the exciton eigenstate as $\ket{X_{n,\mb{Q}}}=\sum_{\mb{k}} \phi_{n\mb{Q}}(\mb{k})b^\dagger_{\mb{Q},\mb{k}}\ket{\text{GS}}$, a superposition of two-particle basis states weighted by an envelope function.
Figure~\ref{fig_Bse}(b) shows the exciton dispersion calculated at a dielectric constant $\epsilon=20$. 
The lowest three exciton bands exhibit a Kagome band structure analogous to that of the tight-binding model [Fig.~\ref{fig_Schematic}d(iii)].
For the lowest band, the exchange interaction produces an approximately $\mb Q$-linear correction near $\gamma$~\cite{yu2014dirac,yu2014valley,qiu2015nonanalyticity}, while leaving all qualitative conclusions unchanged.
The potential $\varphi=\pi/2$ imposes the $\mc{PI}$ symmetry, which enforces a single-particle relation $\varepsilon_{c,\mb{k}}=-\varepsilon_{v,\mb{k}}$ throughout the BZ [Fig.~\ref{fig_Bse}(a)] and hence a centrosymmetric exciton dispersion $\varepsilon^{\text{X}}_{n,\mb{Q}}=\varepsilon^{\text{X}}_{n,-\mb{Q}}$ (see details in the SM~\cite{supp}).

To quantify the exciton topology, we introduce the exciton quantum geometry~\cite{jankowski2025excitonic, yao2008berry,chen2017chiral, blason2020exciton, xie2024theory, davenport2026exciton} (see details in the SM~\cite{supp})
\begin{eqnarray}
    [g_{ab}(\mb{Q})]_n=\braket{\partial_a U_{n,\mb{Q}}|(1-\ket{U_{n,\mb{Q}}}\bra{U_{n,\mb{Q}}})|\partial_b U_{n,\mb{Q}}},
\end{eqnarray}
where $\partial_a \equiv \partial_{Q_a}$, and $\ket{U_{n,\mb{Q}}}$ is the periodic part of the exciton Bloch state, $\ket{U_{n,\mb{Q}}}=e^{-i\mb{Q} \cdot \mb{R}} \ket{X_{n,\mb{Q}}}$, with $\mb{R}$ the COM coordinate. 
The imaginary and real parts of $g_{ab}(\mb{Q})$ give the exciton Berry curvature and quantum metric tensor, respectively.
Figure~\ref{fig_Bse}(b) shows the Berry curvatures of the three exciton Kagome bands, which are concentrated near the Dirac points of the two dispersive bands. 
The corresponding Chern numbers of the three bands, from lowest to highest energy, are $1$, $0$, and $-1$, respectively.
In addition, the exciton flat band exhibits nearly uniform distributions of both the exciton Berry curvature and the quantum metric trace [Fig.~\ref{fig_Bse}(c), cf. Fig.~\ref{fig_Bse}(b)]. 
This flat band arises from the destructive interference in CT-exciton hopping, generated by the kinetic propagation of the electron and hole.

\begin{figure}
\includegraphics[width=0.5\textwidth]{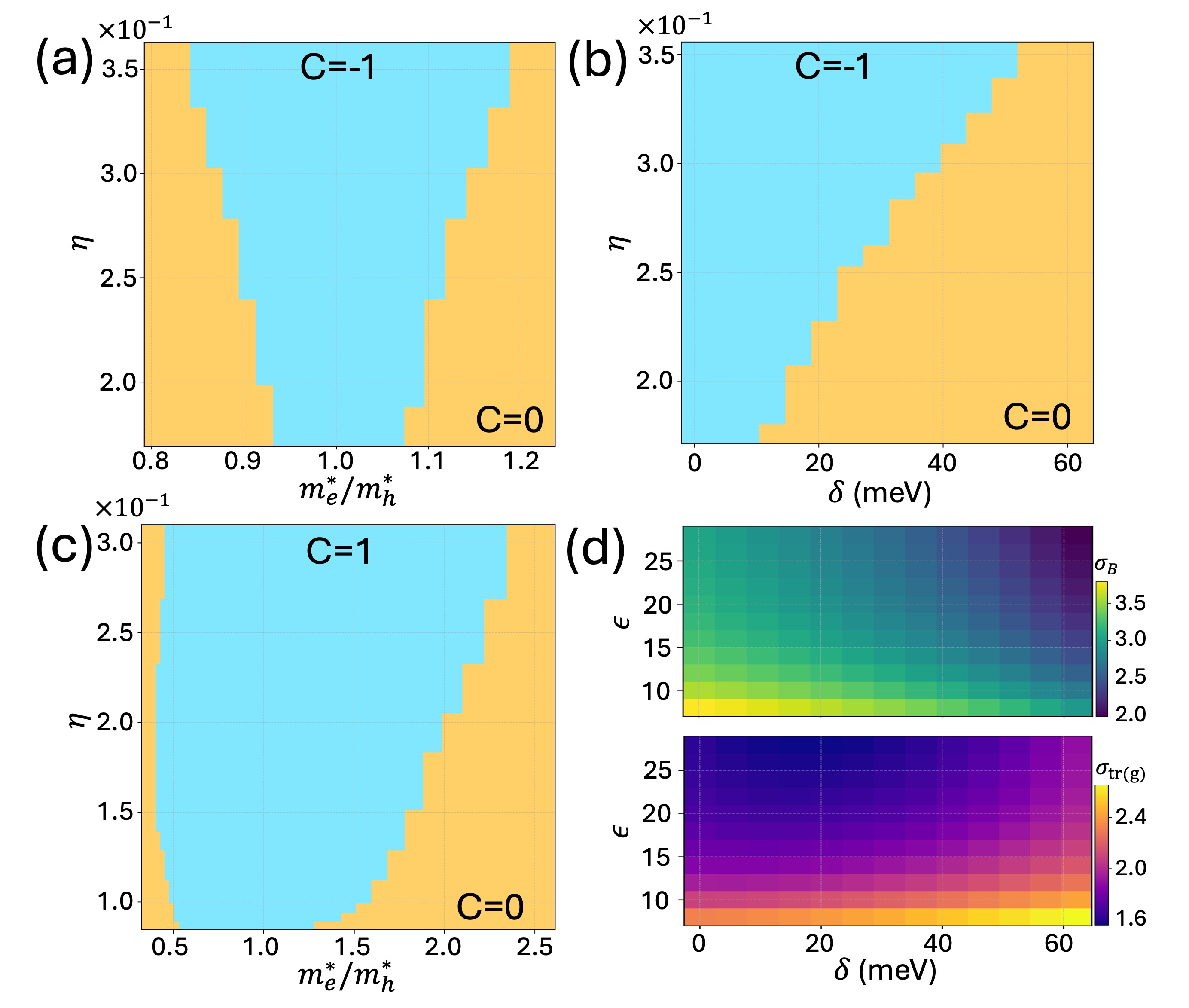}%
\caption{(a) Topological phase diagram of the lowest exciton band as a function of the mass asymmetry $m^*_e/m^*_h$ and flux $\eta$. 
(b) Topological phase diagram of the lowest exciton band as a function of the potential difference $\delta=6\Delta |\cos(\varphi)|$ and flux $\eta$.
(c) Topological phase diagram of the exciton flat band as a function of the mass asymmetry $m^*_e/m^*_h$ and flux $\eta$.
(d) Standard deviations of the exciton Berry curvature (upper) and quantum metric trace (lower) of the exciton flat band as a function of the potential difference $\delta$ and dielectric constant $\epsilon$. 
\label{fig_geometry}}
\end{figure}

\textit{Topology stability under $\mc{PI}$ breaking}---
As we have analyzed, although the Kagome configuration is underpinned by $\mc{PI}$ symmetry, which fixes its bond-centered embedding, the ground-state exciton topology is determined by the competition between $\mc{PI}$ and $\mc{T}$ breaking.
In realistic systems, the absence of $\mc{PI}$ symmetry can be attributed to two origins: the mass and trapping potential asymmetry between the electron and hole.
To quantify the mass asymmetry, we introduce a quadratic term into the Dirac Hamiltonian $H_{\text{Dirac}}\rightarrow H_{\text{Dirac}}+\lambda(\hbar^2\mb{k}^2/2m_e)(I_2+\sigma_z)/2$, where $I_2$ is the identity, $m_e$ is the electron mass, and the dimensionless parameter $\lambda$ describes the strength of the asymmetry. 
The electron mass is thereby reduced (for positive $\lambda$), while the hole mass remains invariant to second order in $k$. 
We parameterize the mass asymmetry by the electron-to-hole mass ratio $m^*_e/m^*_h=1/(1+\lambda m^*/m_e)$, where $m^*\approx \Delta_g/(2v_F^2)$ is the effective mass of the massive Dirac cone.
The trapping potential asymmetry, by contrast, arises when the potential phase $\varphi$ deviates from the $\mc{PI}$-symmetric points $\pm \pi/2$. 
The potential extrema on the two sublattices then become asymmetric, which can be characterized by 
$\delta
=6\Delta |\cos(\varphi)|
$. Thus, by tuning $\varphi$, we use $\delta$ to parameterize $\mc{PI}$ breaking due to trapping potential asymmetry.

To quantify $\mc{T}$ breaking, we use the $k$-space flux $\eta \equiv \hbar^2/(m^*\Delta_g)*A_{BZ}$, a dimensionless parameter equal to the electronic Berry curvature at Dirac point multiplied by the BZ area $A_{BZ}$. 
Because time reversal reverses the valley Berry curvature, this flux provides a natural measure of the effective $\mc{T}$-breaking strength in the single-valley continuum model.
By contrast, the mass asymmetry $m^*_e/m^*_h$ and trapping potential difference $\delta$ are related to the hopping imbalance $\text{Re}(t_c-t_v)$ that breaks $\mc{PI}$ symmetry.

Figures~\ref{fig_geometry}(a) and~\ref{fig_geometry}(b) show the topological phase diagrams of the lowest exciton band as functions of $(\eta,m^*_e/m^*_h)$ and $(\eta,\delta)$, respectively.
Because the topological gaps at $\kappa_{\pm}$ are determined by the competition between $\mc{PI}$ and $\mc{T}$ breaking, a larger flux can tolerate stronger $\mc{PI}$ breaking. 
The band therefore remains topologically nontrivial over broad ranges of mass asymmetry ($\gtrsim 10\%$) and potential difference ($\gtrsim10$ meV).
By contrast, Fig.~\ref{fig_geometry}(c) shows the phase diagram of the exciton flat band [the third band in Fig.~\ref{fig_Bse}(b)].
Its nontrivial topology, induced solely by $\mc{T}$ breaking, persists until orbital localization breaks down.

The uniformity of the quantum geometry of the exciton flat band is also robust. 
Figure~\ref{fig_geometry}(d) shows the standard deviations of the exciton Berry curvature and quantum metric trace as functions of the dielectric constant and potential difference.
The weak momentum dependence and geometric inhomogeneity in projected exciton interactions make the narrow Kagome band a favorable setting for interaction-driven bosonic states.

\begin{figure}
\includegraphics[width=0.5\textwidth]{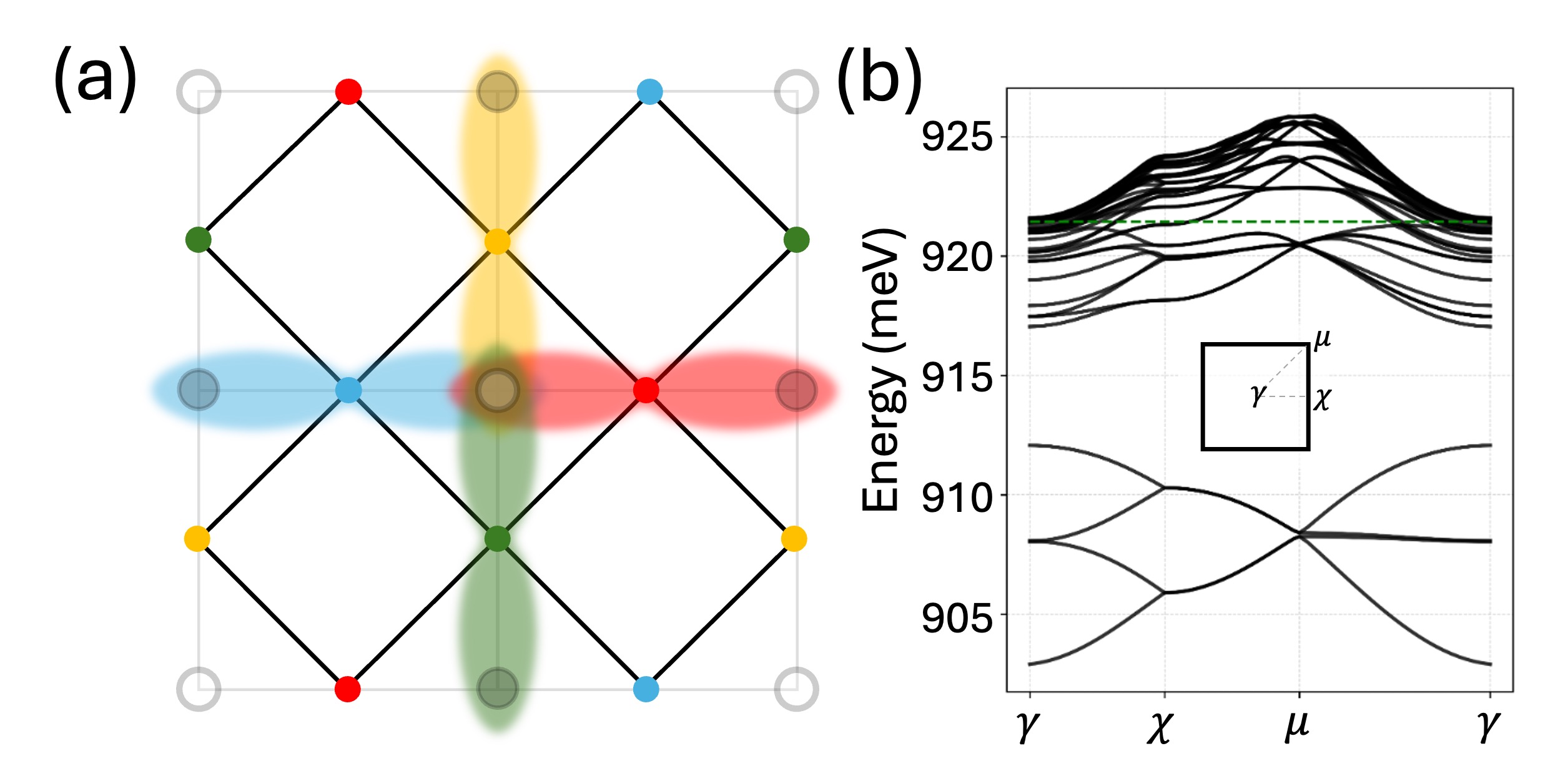}%
\caption{(a) Schematic four-orbital square network of CT excitons formed from a bipartite square network of electrons (solid dots) and holes (hollow dots). 
(b) BSE spectrum of the lowest 50 exciton bands for the model in (a). The green dashed line marks the single-particle gap.}
\label{fig_Discussion}
\end{figure}

\textit{Discussion}---
In summary, our work identifies a general mechanism by which charge-transfer excitons can realize nontrivial crystalline topology.
Because the constituent electron and hole occupy distinct sites, the resulting exciton orbitals can be centered away from either constituent site.
Crystal symmetries then relate these off-site orbitals within each unit cell and organize them into a low-energy multi-orbital manifold.
The emergent lattice of CT excitons is therefore not assumed phenomenologically, but follows from the symmetry-constrained embedding of the composite excitation.
From this perspective, CT excitons provide a tunable route to crystalline topology in composite bound states. 
Importantly, this mechanism is not restricted to the Kagome geometry of CT excitons. 
As a second example, we consider a bipartite square network in which the electron and hole separately occupy distinct fourfold-rotation centers. 
For each fixed hole, there are four distinct relative configurations of electron-hole pairs related by fourfold rotation, forming a four-orbital manifold.
The lowest CT exciton is then expected to reside on the bond connecting the electron and hole, forming a four-orbital square network.
We solve the BSE using the same continuum model as in Eq.~\eqref{eq:sp_hamiltonian}, but replace the triangular potential by a fourfold form
$V(\mb{r})=2\Delta\sum_{j=1,2}\cos(\mb{g}_j \cdot \mb{r} - \varphi)$, with $\varphi=\pi/2$. 
The low-energy spectrum exhibits the corresponding four-orbital exciton manifold [Fig.~\ref{fig_Discussion}(b)], illustrating that the construction is not tied to Kagome geometry. 
Therefore, a systematic extension to other space groups and a topological quantum chemistry for generic composite bound pairs may provide a broader framework for engineering crystalline exciton topology from charge-transfer configurations.

A material realization requires the CT excitons to form the relevant low-energy manifold, with several symmetry-related electron-hole configurations remaining nearly degenerate. Moir\'e systems provide a natural setting for this regime because stacking-dependent potentials can localize electrons and holes at distinct positions within a moir\'e unit cell, as suggested by existing experimental and theoretical studies of transition-metal dichalcogenide (TMD) homo- and heterobilayers and TMD-hBN structures~\cite{li2024imaging, wang2023intercell,naik2022intralayer,maity2026origin,wang2025twist,zhao2021universal,kim2025moire}. 
We also briefly discuss how to connect the low-energy physics of moir\'e heterostructures to our model Hamiltonian~\eqref{eq:sp_hamiltonian} in the SM~\cite{supp}.
The tunability of the moir\'e potential, dielectric environment, and band alignment makes these systems promising platforms for realizing the CT exciton structures proposed here and motivates future material-specific studies to identify favorable regimes and candidate heterostructures.
Experimentally, the nonzero exciton Chern number can manifest through chiral in-gap exciton edge modes, as proposed in previous topological-exciton studies~\cite{gong2017chiral,wu2017topological,chen2017chiral,davenport2024interaction}. Recent spatially resolved luminescence measurements have directly imaged topology-induced excitonic boundary states in two-dimensional molecular crystals~\cite{stein2026topological}, providing an experimental precedent for spectroscopic detection of exciton edge states.

The topological flat band and favorable quantum geometry found here demonstrate the broader potential of CT excitons as a platform for interaction-driven bosonic physics, the resulting many-exciton correlated phases require an explicit treatment of exciton-exciton interactions and remain an important direction for future work. More generally, the central organizing principle is not a particular material or lattice geometry, but the symmetry organization of off-site composite orbitals generated by spatially separated constituents. This mechanism may therefore provide a route to crystalline topological bands in a broader class of composite bound-state systems~\cite{salerno2020interaction}.

\textit{Note added}—Recently, we became aware of an independent work on similar topics~\cite{hwang2026exciton}.

\textit{Acknowledgments}---
The theory development of this work was supported by the U.S. Department of Energy, Office of Basic Energy Sciences, Division of Materials Sciences and Engineering under Contract No. DE-AC02-76SF00515. 
The numerical calculations of excitonic states are supported by the National Science Foundation under Award No. DMR-2339995.
This research used resources of the National Energy Research Scientific Computing Center, a DOE Office of Science User Facility supported by the Office of Science of the U.S. Department of Energy under Contract No. DE-AC02-05CH11231 using NERSC awards BES-ERCAP0037104 and BES-ERCAP0037097.
This work was facilitated through the use of advanced computational, storage, and networking infrastructure provided by the AI-core as well as the Hyak supercomputer system funded by the University of Washington Molecular Engineering Materials Center at the University of Washington (DMR-2308979).

\bibliography{references}

@article{fu2011topological,
  author = {Fu, Liang},
  title = {{Topological Crystalline Insulators}},
  journal = {Phys. Rev. Lett.},
  volume = {106},
  number = {10},
  pages = {106802},
  year = {2011},
  doi = {10.1103/PhysRevLett.106.106802},
  url = {https://doi.org/10.1103/PhysRevLett.106.106802}
}

@article{chernikov2014exciton,
  author = {Chernikov, Alexey and Berkelbach, Timothy C. and Hill, Heather M. and Rigosi, Albert and Li, Yilei and Aslan, Burak and Reichman, David R. and Hybertsen, Mark S. and Heinz, Tony F.},
  title = {{Exciton Binding Energy and Nonhydrogenic Rydberg Series in Monolayer {WS$_2$}}},
  journal = {Phys. Rev. Lett.},
  volume = {113},
  number = {7},
  pages = {076802},
  year = {2014},
  doi = {10.1103/PhysRevLett.113.076802},
  url = {https://doi.org/10.1103/PhysRevLett.113.076802}
}

@article{lozano2025optical,
  author = {Lozano, Mara and Xie, Hong-Yi and Uchoa, Bruno},
  title = {{Optical selection rules of topological excitons in flat bands}},
  journal = {Phys. Rev. B},
  volume = {112},
  number = {23},
  pages = {235417},
  year = {2025},
  doi = {10.1103/t9my-h5gy},
  url = {https://doi.org/10.1103/t9my-h5gy}
}

@article{froese2025topological,
  author = {Froese, Paul and Neupert, Titus and Wagner, Glenn},
  title = {{Topological excitons in moir{\'e} {MoTe$_2$/WSe$_2$} heterobilayers}},
  journal = {Phys. Rev. Res.},
  volume = {7},
  number = {2},
  pages = {023047},
  year = {2025},
  doi = {10.1103/PhysRevResearch.7.023047},
  url = {https://doi.org/10.1103/PhysRevResearch.7.023047}
}

@article{kwan2021exciton,
  author = {Kwan, Yves H. and Hu, Yichen and Simon, Steven H. and Parameswaran, S. A.},
  title = {{Exciton Band Topology in Spontaneous Quantum Anomalous Hall Insulators: Applications to Twisted Bilayer Graphene}},
  journal = {Phys. Rev. Lett.},
  volume = {126},
  number = {13},
  pages = {137601},
  year = {2021},
  doi = {10.1103/PhysRevLett.126.137601},
  url = {https://doi.org/10.1103/PhysRevLett.126.137601}
}

@article{xie2024long,
  author = {Xie, Ming and Hafezi, Mohammad and Das Sarma, Sankar},
  title = {{Long-Lived Topological Flatband Excitons in Semiconductor Moir{\'e} Heterostructures: A Bosonic Kane-Mele Model Platform}},
  journal = {Phys. Rev. Lett.},
  volume = {133},
  number = {13},
  pages = {136403},
  year = {2024},
  doi = {10.1103/PhysRevLett.133.136403},
  url = {https://doi.org/10.1103/PhysRevLett.133.136403}
}

@article{wu2017topological,
  author = {Wu, Fengcheng and Lovorn, Timothy and MacDonald, A. H.},
  title = {{Topological Exciton Bands in Moir{\'e} Heterojunctions}},
  journal = {Phys. Rev. Lett.},
  volume = {118},
  number = {14},
  pages = {147401},
  year = {2017},
  doi = {10.1103/PhysRevLett.118.147401},
  url = {https://doi.org/10.1103/PhysRevLett.118.147401}
}

@article{davenport2024interaction,
  author = {Davenport, Henry and Knolle, Johannes and Schindler, Frank},
  title = {{Interaction-Induced Crystalline Topology of Excitons}},
  journal = {Phys. Rev. Lett.},
  volume = {133},
  number = {17},
  pages = {176601},
  year = {2024},
  doi = {10.1103/PhysRevLett.133.176601},
  url = {https://doi.org/10.1103/PhysRevLett.133.176601}
}

@article{jankowski2025excitonic,
  author = {Jankowski, Wojciech J. and Thompson, Joshua J. P. and Monserrat, Bartomeu and Slager, Robert-Jan},
  title = {{Excitonic topology and quantum geometry in organic semiconductors}},
  journal = {Nat. Commun.},
  volume = {16},
  number = {1},
  pages = {4661},
  year = {2025},
  doi = {10.1038/s41467-025-59257-5},
  url = {https://doi.org/10.1038/s41467-025-59257-5}
}

@article{davenport2026exciton,
    author = {Davenport, Henry and Knolle, Johannes and Schindler, Frank},
    title = {{Exciton Berryology}},
    journal = {Phys. Rev. B},
    volume = {113},
    number = {4},
    pages = {045125},
    year = {2026},
    doi = {10.1103/jtgq-vc7n},
    url = {https://doi.org/10.1103/jtgq-vc7n}
  }

@article{zhang2025engineering,
  author = {Zhang, Na and Yao, Wang and Yu, Hongyi},
  title = {{Engineering topological exciton structures in two-dimensional semiconductors by a periodic electrostatic potential}},
  journal = {Phys. Rev. B},
  volume = {112},
  number = {16},
  pages = {165406},
  year = {2025},
  doi = {10.1103/xhgh-mpky},
  url = {https://doi.org/10.1103/xhgh-mpky}
}

@article{zheng2025forster,
  author = {Zheng, Huiyuan and Li, Ci and Yu, Hongyi and Yao, Wang},
  title = {{F{\"o}rster valley-orbit coupling and topological lattice of hybrid moir{\'e} excitons}},
  journal = {Commun. Phys.},
  volume = {8},
  number = {1},
  pages = {193},
  year = {2025},
  doi = {10.1038/s42005-025-02114-0},
  url = {https://doi.org/10.1038/s42005-025-02114-0}
}

@article{naik2022intralayer,
  author = {Naik, Mit H. and Regan, Emma C. and Zhang, Zuocheng and Chan, Yang-Hao and Li, Zhenglu and Wang, Danqing and Yoon, Yoseob and Ong, Chin Shen and Zhao, Wenyu and Zhao, Sihan and Utama, M. Iqbal Bakti and Gao, Beini and Wei, Xin and Sayyad, Mohammed and Yumigeta, Kentaro and Watanabe, Kenji and Taniguchi, Takashi and Tongay, Sefaattin and {da Jornada}, Felipe H. and Wang, Feng and Louie, Steven G.},
  title = {{Intralayer charge-transfer moir{\'e} excitons in van der Waals superlattices}},
  journal = {Nature},
  volume = {609},
  number = {7925},
  pages = {52--57},
  year = {2022},
  doi = {10.1038/s41586-022-04991-9},
  url = {https://doi.org/10.1038/s41586-022-04991-9}
}

@article{zeng2022strong,
  author = {Zeng, Yongxin and MacDonald, Allan H.},
  title = {{Strong modulation limit of excitons and trions in moir{\'e} materials}},
  journal = {Phys. Rev. B},
  volume = {106},
  number = {3},
  pages = {035115},
  year = {2022},
  doi = {10.1103/PhysRevB.106.035115},
  url = {https://doi.org/10.1103/PhysRevB.106.035115}
}

@article{shradha20262d,
  author = {Shradha, S and Rosati, R and Lamsaadi, H and Picker, J and Paradisanos, I and Hossain, Md T and Krelle, L and Oswald, L F and Engel, N and Markina, D I and Watanabe, K and Taniguchi, T and Sahoo, P K and Lombez, L and Marie, X and Renucci, P and Paillard, V and Poumirol, J-M and Turchanin, A and Malic, E and Urbaszek, B},
  title = {{2D excitonics with atomically thin lateral heterostructures}},
  journal = {Rep. Prog. Phys.},
  volume = {89},
  number = {4},
  pages = {046501},
  year = {2026},
  doi = {10.1088/1361-6633/ae530c},
  url = {https://doi.org/10.1088/1361-6633/ae530c}
}

@article{rosati2023interface,
  author = {Rosati, Roberto and Paradisanos, Ioannis and Huang, Libai and Gan, Ziyang and George, Antony and Watanabe, Kenji and Taniguchi, Takashi and Lombez, Laurent and Renucci, Pierre and Turchanin, Andrey and Urbaszek, Bernhard and Malic, Ermin},
  title = {{Interface engineering of charge-transfer excitons in 2D lateral heterostructures}},
  journal = {Nat. Commun.},
  volume = {14},
  number = {1},
  pages = {2438},
  year = {2023},
  doi = {10.1038/s41467-023-37889-9},
  url = {https://doi.org/10.1038/s41467-023-37889-9}
}

@article{wang2025twist,
  author = {Wang, Renqi and Chang, Kai and Duan, Wenhui and Xu, Yong and Tang, Peizhe},
  title = {{Twist-Angle-Dependent Valley Polarization of Intralayer Moir{\'e} Excitons in van der Waals Superlattices}},
  journal = {Phys. Rev. Lett.},
  volume = {134},
  number = {2},
  pages = {026904},
  year = {2025},
  doi = {10.1103/PhysRevLett.134.026904},
  url = {https://doi.org/10.1103/PhysRevLett.134.026904}
}

@article{yu2014dirac,
  author = {Yu, Hongyi and Liu, Gui-Bin and Gong, Pu and Xu, Xiaodong and Yao, Wang},
  title = {{Dirac cones and Dirac saddle points of bright excitons in monolayer transition metal dichalcogenides}},
  journal = {Nat. Commun.},
  volume = {5},
  number = {1},
  pages = {3876},
  year = {2014},
  doi = {10.1038/ncomms4876},
  url = {https://doi.org/10.1038/ncomms4876}
}

@article{yu2014valley,
  author = {Yu, T. and Wu, M. W.},
  title = {{Valley depolarization due to intervalley and intravalley electron-hole exchange interactions in monolayer {MoS$_2$}}},
  journal = {Phys. Rev. B},
  volume = {89},
  number = {20},
  pages = {205303},
  year = {2014},
  doi = {10.1103/PhysRevB.89.205303},
  url = {https://doi.org/10.1103/PhysRevB.89.205303}
}

@article{qiu2015nonanalyticity,
  author = {Qiu, Diana Y. and Cao, Ting and Louie, Steven G.},
  title = {{Nonanalyticity, Valley Quantum Phases, and Lightlike Exciton Dispersion in Monolayer Transition Metal Dichalcogenides: Theory and First-Principles Calculations}},
  journal = {Phys. Rev. Lett.},
  volume = {115},
  number = {17},
  pages = {176801},
  year = {2015},
  doi = {10.1103/PhysRevLett.115.176801},
  url = {https://doi.org/10.1103/PhysRevLett.115.176801}
}

@Misc{supp,
  howpublished	= "See Supplemental Material for (1) Material realization of the single-particle model; (2) BSE formalism in the moir\'e band basis; (3) Strong-modulation limit and CT tight binding model; (4) Particle-hole and combined particle-hole-inversion symmetry; (5) Exciton Berry connection, curvature and Wilson-line evaluation, which includes Refs.[26-28]."
}

@article{yang2026giant,
  title = {Giant and helical exciton dipole from Berry curvature in flat Chern bands},
  author = {Yang, Kaijie and Zheng, Huiyuan and Xu, Xiaodong and Xiao, Di and Cao, Ting},
  journal = {Phys. Rev. Lett.},
  volume = {137},
  issue = {12},
  pages = {126602},
  numpages = {8},
  year = {2026},
  month = {Sep},
  publisher = {American Physical Society},
  doi = {10.1103/5w7l-8mcv},
  url = {https://link.aps.org/doi/10.1103/5w7l-8mcv}
}

@article{li2024imaging,
  author = {Li, Hongyuan and Xiang, Ziyu and Naik, Mit H. and Kim, Woochang and Li, Zhenglu and Sailus, Renee and Banerjee, Rounak and Taniguchi, Takashi and Watanabe, Kenji and Tongay, Sefaattin and Zettl, Alex and da Jornada, Felipe H. and Louie, Steven G. and Crommie, Michael F. and Wang, Feng},
  title = {{Imaging moir{\'e} excited states with photocurrent tunnelling microscopy}},
  journal = {Nat. Mater.},
  volume = {23},
  number = {5},
  pages = {633--638},
  year = {2024},
  doi = {10.1038/s41563-023-01753-4},
  url = {https://doi.org/10.1038/s41563-023-01753-4}
}

@article{wang2023intercell,
  author = {Wang, Xi and Zhang, Xiaowei and Zhu, Jiayi and Park, Heonjoon and Wang, Yingqi and Wang, Chong and Holtzmann, William G. and Taniguchi, Takashi and Watanabe, Kenji and Yan, Jiaqiang and Gamelin, Daniel R. and Yao, Wang and Xiao, Di and Cao, Ting and Xu, Xiaodong},
  title = {{Intercell moir{\'e} exciton complexes in electron lattices}},
  journal = {Nat. Mater.},
  volume = {22},
  number = {5},
  pages = {599--604},
  year = {2023},
  doi = {10.1038/s41563-023-01496-2},
  url = {https://doi.org/10.1038/s41563-023-01496-2}
}

@article{choi1964charge,
  author = {Choi, Sang-Il and Jortner, Joshua and Rice, Stuart A. and Silbey, Robert},
  title = {{Charge-Transfer Exciton States in Aromatic Molecular Crystals}},
  journal = {J. Chem. Phys.},
  volume = {41},
  number = {11},
  pages = {3294--3306},
  year = {1964},
  doi = {10.1063/1.1725728},
  url = {https://doi.org/10.1063/1.1725728}
}

@article{pereira2019electroabsorption,
  author = {Pereira, Daniel de Sa and Menelaou, Christopher and Danos, Andrew and Marian, Christel and Monkman, Andrew P.},
  title = {{Electroabsorption Spectroscopy as a Tool for Probing Charge Transfer and State Mixing in Thermally Activated Delayed Fluorescence Emitters}},
  journal = {J. Phys. Chem. Lett.},
  volume = {10},
  number = {12},
  pages = {3205--3211},
  year = {2019},
  doi = {10.1021/acs.jpclett.9b00999},
  url = {https://doi.org/10.1021/acs.jpclett.9b00999}
}

@book{haug2009quantum,
    author = {Haug, Hartmut and Koch, Stephan W.},
    title = {Quantum Theory of the Optical and Electronic Properties of Semiconductors},
    edition = {5},
    publisher = {World Scientific},
    address = {Singapore},
    year = {2009},
    doi = {10.1142/7184},
    isbn = {9789812838834},
    url = {https://doi.org/10.1142/7184}
  }

@article{zhao2021universal,
  author = {Zhao, Pei and Xiao, Chengxin and Yao, Wang},
  title = {{Universal superlattice potential for 2D materials from twisted interface inside h-BN substrate}},
  journal = {npj 2D Mater. Appl.},
  volume = {5},
  number = {1},
  pages = {38},
  year = {2021},
  doi = {10.1038/s41699-021-00221-4},
  url = {https://doi.org/10.1038/s41699-021-00221-4}
}

@article{kim2025moire,
  author = {Kim, Dong Seob and Xiao, Chengxin and Dominguez, Roy C. and Liu, Zhida and Abudayyeh, Hamza and Lee, Kyoungpyo and Mayorga-Luna, Rigo and Kim, Hyunsue and Watanabe, Kenji and Taniguchi, Takashi and Shih, Chih-Kang and Miyahara, Yoichi and Yao, Wang and Li, Xiaoqin},
  title = {{Moir{\'e} ferroelectricity modulates light emission from a semiconductor monolayer}},
  journal = {Sci. Adv.},
  volume = {11},
  number = {19},
  pages = {eadt7789},
  year = {2025},
  doi = {10.1126/sciadv.adt7789},
  url = {https://doi.org/10.1126/sciadv.adt7789}
}

@article{chen2017chiral,
  author = {Chen, Ke and Shindou, Ryuichi},
  title = {{Chiral topological excitons in a Chern band insulator}},
  journal = {Phys. Rev. B},
  volume = {96},
  number = {16},
  pages = {161101},
  year = {2017},
  doi = {10.1103/physrevb.96.161101},
  url = {https://doi.org/10.1103/physrevb.96.161101}
}

@article{blason2020exciton,
  author = {Blason, Andrea and Fabrizio, Michele},
  title = {{Exciton topology and condensation in a model quantum spin Hall insulator}},
  journal = {Phys. Rev. B},
  volume = {102},
  number = {3},
  pages = {035146},
  year = {2020},
  doi = {10.1103/physrevb.102.035146},
  url = {https://doi.org/10.1103/physrevb.102.035146}
}

@article{xie2024theory,
  author = {Xie, Hong-Yi and Ghaemi, Pouyan and Mitrano, Matteo and Uchoa, Bruno},
  title = {{Theory of topological exciton insulators and condensates in flat Chern bands}},
  journal = {Proc. Natl. Acad. Sci. U.S.A.},
  volume = {121},
  number = {35},
  pages = {e2401644121},
  year = {2024},
  doi = {10.1073/pnas.2401644121},
  url = {https://doi.org/10.1073/pnas.2401644121}
}

@article{fang2012bulk,
  author = {Fang, Chen and Gilbert, Matthew J. and Bernevig, B. Andrei},
  title = {{Bulk topological invariants in noninteracting point group symmetric insulators}},
  journal = {Phys. Rev. B},
  volume = {86},
  number = {11},
  pages = {115112},
  year = {2012},
  doi = {10.1103/physrevb.86.115112},
  url = {https://doi.org/10.1103/physrevb.86.115112}
}

@article{bradlyn2017topological,
  author = {Bradlyn, Barry and Elcoro, L. and Cano, Jennifer and Vergniory, M. G. and Wang, Zhijun and Felser, C. and Aroyo, M. I. and Bernevig, B. Andrei},
  title = {{Topological quantum chemistry}},
  journal = {Nature},
  volume = {547},
  number = {7663},
  pages = {298--305},
  year = {2017},
  doi = {10.1038/nature23268},
  url = {https://doi.org/10.1038/nature23268}
}

@article{salerno2020interaction,
  author = {Salerno, G. and Palumbo, G. and Goldman, N. and Di Liberto, M.},
  title = {{Interaction-induced lattices for bound states: Designing flat bands, quantized pumps, and higher-order topological insulators for doublons}},
  journal = {Phys. Rev. Res.},
  volume = {2},
  number = {1},
  pages = {013348},
  year = {2020},
  doi = {10.1103/physrevresearch.2.013348},
  url = {https://doi.org/10.1103/physrevresearch.2.013348}
}

@article{chiu2016classification,
  author = {Chiu, Ching-Kai and Teo, Jeffrey C. Y. and Schnyder, Andreas P. and Ryu, Shinsei},
  title = {{Classification of topological quantum matter with symmetries}},
  journal = {Rev. Mod. Phys.},
  volume = {88},
  number = {3},
  pages = {035005},
  year = {2016},
  doi = {10.1103/revmodphys.88.035005},
  url = {https://doi.org/10.1103/revmodphys.88.035005}
}

@article{xiao2012coupled,
  author = {Xiao, Di and Liu, Gui-Bin and Feng, Wanxiang and Xu, Xiaodong and Yao, Wang},
  title = {{Coupled Spin and Valley Physics in Monolayers of {MoS$_2$} and Other Group-{VI} Dichalcogenides}},
  journal = {Phys. Rev. Lett.},
  volume = {108},
  number = {19},
  pages = {196802},
  year = {2012},
  doi = {10.1103/physrevlett.108.196802},
  url = {https://doi.org/10.1103/physrevlett.108.196802}
}

@article{maity2026origin,
  author = {Maity, Indrajit and Lischner, Johannes and Mostofi, Arash A. and Rubio, {\'A}ngel},
  title = {{Origin of Trapped Intralayer Wannier and Charge-Transfer Excitons in Moir{\'e} Materials}},
  journal = {Nano Lett.},
  volume = {26},
  number = {4},
  pages = {1349--1356},
  year = {2026},
  doi = {10.1021/acs.nanolett.5c05352},
  url = {https://doi.org/10.1021/acs.nanolett.5c05352}
}

@article{yao2008berry,
  author = {Yao, Wang and Niu, Qian},
  title = {{Berry Phase Effect on the Exciton Transport and on the Exciton Bose-Einstein Condensate}},
  journal = {Phys. Rev. Lett.},
  volume = {101},
  number = {10},
  pages = {106401},
  year = {2008},
  doi = {10.1103/physrevlett.101.106401},
  url = {https://doi.org/10.1103/physrevlett.101.106401}
}

@article{gong2017chiral,
  author = {Gong, Z. R. and Luo, W. Z. and Jiang, Z. F. and Fu, H. C.},
  title = {{Chiral topological excitons in the monolayer transition metal dichalcogenides}},
  journal = {Sci. Rep.},
  volume = {7},
  number = {1},
  pages = {42390},
  year = {2017},
  doi = {10.1038/srep42390},
  url = {https://doi.org/10.1038/srep42390}
}

@Misc{stein2026topological,
  archiveprefix = {arXiv},
  arxivid = {2610.09149},
  title = {{Topological Frenkel excitons in 2D molecular crystals}},
  author = {Stein, Quentin and Nayak, Sidhartha and Jiang, Song and Kuhlke, Jakob and Romeo, Michelangelo and Frederiksen, Thomas and Scheurer, Fabrice and Kaiser, Katharina and Neuman, Tomas and Schull, Guillaume},
  eprint = {2610.09149}
}

@book{tasaki2020physics,
  author = {Tasaki, Hal},
  title = {Physics and Mathematics of Quantum Many-Body Systems},
  publisher = {Springer},
  address = {Cham},
  series = {Graduate Texts in Physics},
  edition = {1},
  year = {2020},
  isbn = {978-3-030-41264-7},
  doi = {10.1007/978-3-030-41265-4},
  url = {https://doi.org/10.1007/978-3-030-41265-4}
}

@Misc{hwang2026exciton,
  archiveprefix = {arXiv},
  arxivid = {2608.07372},
  title = {{Exciton Alchemy: Chern Excitons from Trivial Bands}},
  author = {Hwang, Yoonseok and Davenport, Henry and Schindler, Frank},
  eprint = {2608.07372}
}


\clearpage
\pagebreak
\widetext
\begin{center}
	{\large{\bf Supplementary Materials for ``Topological Charge-Transfer Excitons" }}
\end{center}

\setcounter{equation}{0}
\setcounter{figure}{0}
\setcounter{secnumdepth}{2}
\setcounter{page}{1}
\setcounter{section}{0}
\renewcommand{\theequation}{S\arabic{equation}}
\renewcommand{\thefigure}{S\arabic{figure}}
\renewcommand{\thetable}{S\arabic{table}}

\tableofcontents
\bigskip

\section{Material Realization of the Single-Particle Model\label{section:sp_Hamiltonian}}

\begin{figure}
\includegraphics[width=0.7\textwidth]{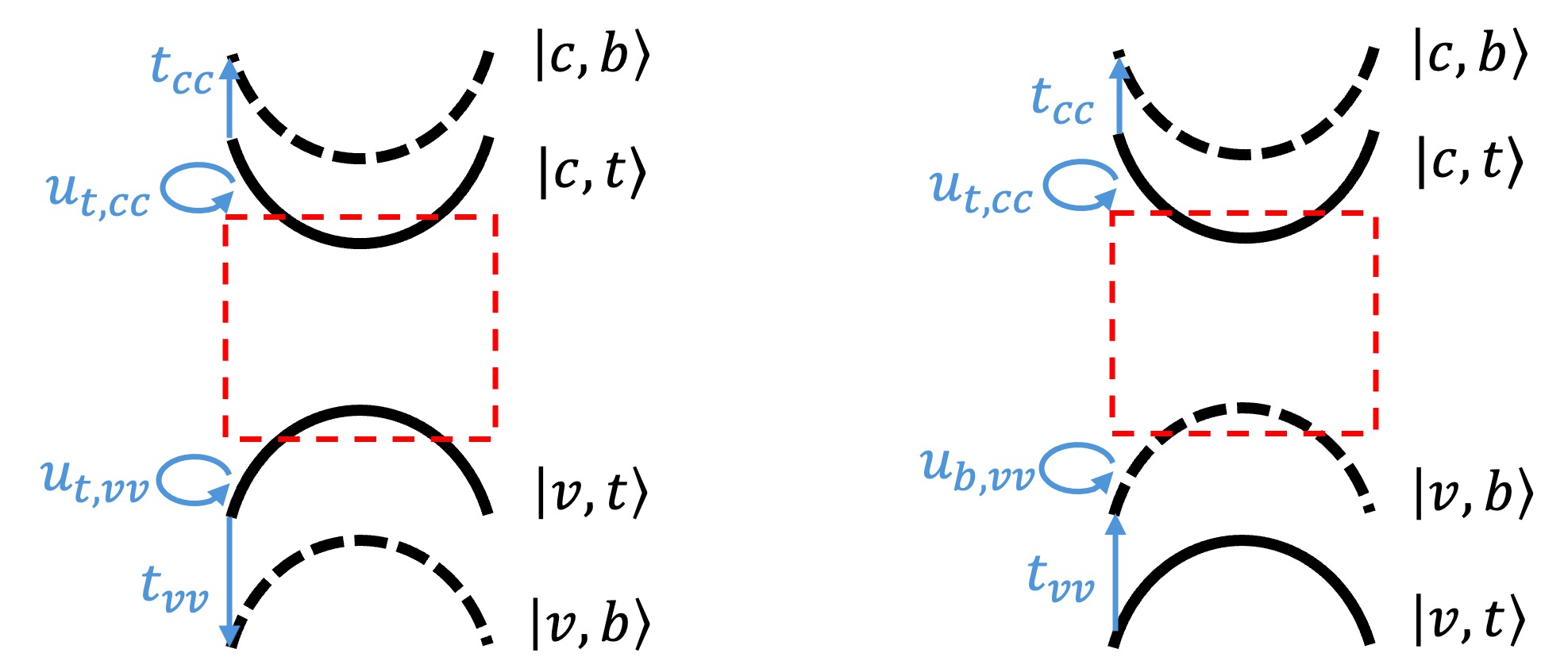}%
\caption{Schematic band alignments of a H-stacking moir\'e bilayer for type-I (left panel) and type-II (right panel) alignment. 
Solid (dashed) curves represent band edges from top (bottom) layer.
Red dashed square represents the projected subspaces. Only the leading coupling terms are displayed.
\label{figS_bandaligment}}
\end{figure}

In this section we discuss how to achieve the picture discussed in the main text from low-energy bands in moir\'e superlattices.
The continuum Hamiltonian in the main text describes a massive Dirac cone modulated by an adjacent moir\'e potential, which can be modeled from a monolayer transition metal dichalcogenide (TMD) placed on twisted hexagonal boron nitrides~\cite{zhao2021universal,kim2025moire}.
The single massive Dirac cone comes from the valley-polarized states, in which the valley selection can be produced by Zeeman splitting, magnetic proximity, spontaneous valley polarization, or valley-selective optical preparation.
In addition, the single-particle Hamiltonian can also be obtained as the low-energy projection of a moir\'e bilayer continuum model.
Consider a generic heterobilayer, let $l=t,b$ label the layers, and let the two orbital components of each layer form a massive Dirac cone $h_l(\mb k_l)$. In a fixed-spin sector, the four-orbital continuum Hamiltonian is

\begin{equation}
\mc H(\mb k,\mb r)
=
\begin{pmatrix}
h_t(\mb k_t)+U_t(\mb r)&T^\dagger(\mb r)\\
T(\mb r)&h_b(\mb k_b)+U_b(\mb r)
\end{pmatrix},
\end{equation}
where the layer offsets are included in $h_l$, while $U_l(\mb r)$ and $T(\mb r)$ are the moir\'e-periodic intralayer modulation and interlayer tunneling. We separate the layer-diagonal kinetic Hamiltonian $\mc H^{(0)}=\operatorname{diag}\{h_t,h_b\}$ from the moir\'e perturbation $W=\mc H-\mc H^{(0)}$, and
use $\ket{\alpha,l} (\alpha=c,v)$ to denote an unperturbed band-edge state of the isolated layer $l$.
The schematic band alignments are shown in Fig.~\ref{figS_bandaligment}.

Let $P$ project onto two isolated band-edge states of $\mc H^{(0)}(\mb 0)$ and let $Q=1-P$ contain the two remote states. To second order in the moir\'e perturbation, Schrieffer-Wolff projection gives

\begin{equation}
H_{\rm eff}(E_0)
=
H_{PP}
+H_{PQ}(E_0-H_{QQ})^{-1}H_{QP},
\end{equation}
where $H_{XY}=X\mc H Y$. 
This projection is controlled channel by channel. For every
retained state $|p\rangle\in P$ and remote state $|q\rangle\in Q$ connected
by the moir\'e perturbation, one requires
\begin{equation}
\frac{|\langle q|U_l(\mb r)|p\rangle|}
     {|E_p^{(0)}-E_q^{(0)}|}
\ll 1,
\qquad
\frac{|\langle q|T^{(\dagger)}(\mb r)|p\rangle|}
     {|E_p^{(0)}-E_q^{(0)}|}
\ll 1,
\end{equation}
where the first and second conditions apply to same-layer and interlayer
couplings, respectively. Thus, both the coupling matrix element and its
energy denominator depend on the chosen $P$ and $Q$ subspaces. 
The two retained states may be the conduction- and valence-edge states of one layer in a type-I alignment or states from opposite layers in a type-II alignment.
The $\mb k$-dependent off-diagonal term of the projected Hamiltonian is fixed by the $C_3$ representations of the retained states. If their $C_3$ angular momenta differ by $\pm1$ modulo three, the leading allowed coupling is proportional to $k_\mp=k_x\mp ik_y$, giving $\hbar v_F\mb k\cdot\bs{\sigma}$ after a basis-phase choice.

To make the potential reduction explicit, define $u_{l,\alpha\beta}=\langle\alpha,l|U_l|\beta,l\rangle$ and $t_{\alpha\beta}=\langle\alpha,b|T|\beta,t\rangle$, where $\alpha,\beta=c,v$. The projected moir\'e potential is

\begin{equation}
V_{\rm eff}(\mb r)
=
\begin{pmatrix}
V_c(\mb r)&V_{cv}(\mb r)\\
V_{cv}^*(\mb r)&V_v(\mb r)
\end{pmatrix}.
\end{equation}
For H stacking, spin conservation and the $C_3$ orbital representations determine which $t_{\alpha\beta}$ channels are allowed.
For type-I alignment (left panel in Fig.~\ref{figS_bandaligment}), take $P=\{|c,t\rangle,|v,t\rangle\}$ and $Q=\{|c,b\rangle,|v,b\rangle\}$. The Dirac term is inherited directly from $h_t$, while the off-diagonal moir\'e potential is

\begin{equation}
V_{cv}^{\rm I}
=
u_{t,cv}
+\lambda_{cb}t_{cc}^*t_{cv}
+\lambda_{vb}t_{vc}^*t_{vv},
\end{equation}
where $\lambda_{\alpha b}=(E_0-E_{\alpha b})^{-1}$. 
The diagonal terms are
\begin{equation}
\begin{aligned}
V_c^{\rm I}
&=u_{t,cc}+\lambda_{cb}|t_{cc}|^2+\lambda_{vb}|t_{vc}|^2,\\
V_v^{\rm I}
&=u_{t,vv}+\lambda_{cb}|t_{cv}|^2+\lambda_{vb}|t_{vv}|^2.
\end{aligned}
\end{equation}
The dominant intralayer matrix element is expected to be band diagonal. For a smooth electrostatic modulation, $U_l(\mb r)$ is approximately diagonal in the band-edge orbital basis, so that $|u_{l,cv}|\ll |u_{l,cc}|$.
In particular, $u_{l,cv}=0$ at the band edge when $U_l$ is proportional to the identity in orbital space. 
Similarly, spin conservation and the $C_3$ orbital representations generally favor band-preserving tunneling, giving $|t_{cv}|\ll |t_{cc}|$,
with $t_{cv}=0$ at leading order when the cross-band channel is symmetry forbidden. 

For type-II alignment (right panel in Fig.~\ref{figS_bandaligment}), take $P=\{|c,t\rangle,|v,b\rangle\}$ and $Q=\{|v,t\rangle,|c,b\rangle\}$. The leading static off-diagonal potential is
\begin{equation}
V_{cv}^{\rm II}
=
t_{vc}^*
+\lambda_{vt}u_{t,cv}t_{vv}^*
+\lambda_{cb}t_{cc}^*u_{b,cv}.
\end{equation}
For band-diagonal $U_l$, this reduces to $V_{cv}^{\rm II}=t_{vc}^*$ and is negligible when the direct cross-band tunneling is suppressed by spin or orbital selection rules. The corresponding diagonal terms are

\begin{equation}
\begin{aligned}
V_c^{\rm II}
&=u_{t,cc}+\lambda_{vt}|u_{t,cv}|^2+\lambda_{cb}|t_{cc}|^2,\\
V_v^{\rm II}
&=u_{b,vv}+\lambda_{vt}|t_{vv}|^2+\lambda_{cb}|u_{b,cv}|^2.
\end{aligned}
\end{equation}
The band-preserving channels $t_{cc}$ and $t_{vv}$ may nevertheless generate the linear coupling

\begin{equation}
\Gamma_{\rm II}(\mb k,\mb r)
=
\lambda_{vt}d_t(\mb k)t_{vv}^*(\mb r)
+\lambda_{cb}t_{cc}^*(\mb r)d_b(\mb k),
\end{equation}
where $d_l(\mb k)=\langle c,l|h_l(\mb k)-h_l(\mb 0)|v,l\rangle$. Its uniform leading component gives the effective Dirac term; the remaining spatially modulated velocity and gradient terms are neglected. Hence the type-II realization requires at least one band-preserving tunneling channel to survive while the direct cross-band channel remains small.
H-stacked moir\'e heterobilayers remain qualitative candidates for realizing the CT-exciton configuration and a type-II realization of the specific two-band Dirac model requires appreciable symmetry-allowed band-preserving tunneling.

With the two-band model simplified from full moir\'e bilayer continuum model, the diagonal potential terms $V_c$ and $V_v$ then contain the directly projected band-edge modulation together with the second-order tunneling self-energy $W_{PQ}(E_0-H_{QQ}^{(0)})^{-1}W_{QP}$. Because the tunneling amplitudes are moir\'e-periodic, their quadratic self-energies are also moir\'e-periodic. Retaining the first $C_3$-related reciprocal-lattice shell gives

\begin{equation}
V_{c/v}(\mb r)
=
2\Delta_{c/v}\sum_{j=1}^3
\cos(\mb g_j\cdot\mb r-\varphi_{c/v}).
\end{equation}
Neglecting higher-order momentum terms, spatially modulated velocities, and the small off-diagonal moir\'e potential therefore yields

\begin{equation}
H_0
=
\hbar v_F \mb k \cdot \bs \sigma
+\frac{\Delta_g}{2}\sigma_z
+\operatorname{diag}\{V_c(\mb r),V_v(\mb r)\}.
\label{eq:continuum_cv}
\end{equation}
This is the general single-particle model used throughout the supplement.
Based on the above discussion, we briefly discuss what determines the mass term. Its value is dictated by the underlying electronic structure: for a monolayer TMD it is primarily set by the material-dependent conduction--valence gap of the active spin/valley sector, whereas in a heterobilayer it is determined by the relative band alignment and interlayer hybridization. In particular, for a type-II alignment, where the retained conduction and valence states reside predominantly in different layers, their relative energy and hence $\Delta_g$ can be tuned by an out-of-plane displacement field.

\section{BSE Formalism in the Moir\'e Band Basis\label{section:BSE}}

In this section, we derive the BSE formalism in the moir\'e band basis. 
The single-particle model used is Eq.~\eqref{eq:continuum_cv}.
Let $e^\dagger_{\alpha,\mb p}$ create an electron in moir\'e band $\alpha$ and crystal momentum $\mb p\in{\rm BZ}$. The underlying electronic orbital has an $A/B$ sublattice label $\mu=A,B$ and a plane-wave moir\'e reciprocal vector $\mb G$. The Bloch-band operator and its inverse are

\begin{equation}
\begin{aligned}
e^\dagger_{\alpha,\mb p}
&=\sum_{\mu,\mb G}
U_{\alpha,\mu\mb G}(\mb p)
a^\dagger_{\mu,\mb p+\mb G},\\
a^\dagger_{\mu,\mb p+\mb G}
&=\sum_\alpha
U^*_{\alpha,\mu\mb G}(\mb p)
e^\dagger_{\alpha,\mb p}.
\end{aligned}
\end{equation}
Using the exciton basis $b^\dagger_{\mb Q,\mb k}=e^\dagger_{c,\mb k+\mb Q}e_{v,\mb k}$, with COM momentum $\mb Q$ and relative momentum $\mb k$, a neutral exciton with COM momentum $\mb Q$ is expanded as

\begin{equation}
|X_{n,\mb Q}\rangle
=\sum_{c v\mb k}
\phi_{n\mb Q,cv}(\mb k)
b^\dagger_{\mb Q,\mb k;cv}|{\rm GS}\rangle .
\end{equation}
For most formulas in the main text, only the lowest isolated conduction band and highest isolated valence band are retained, so the indices $c,v$ can be suppressed.

The BSE is the eigenvalue problem

\begin{equation}
\sum_{cv\mb k}
H_{\mb Q;c'v',cv}(\mb k',\mb k)
\phi_{n\mb Q,cv}(\mb k)
=E_n(\mb Q)\phi_{n\mb Q,c'v'}(\mb k').
\end{equation}
The one-particle part is

\begin{equation}
H^{(1)}_{\mb Q;c'v',cv}(\mb k',\mb k)
=\delta_{c'c}\delta_{v'v}\delta_{\mb k'\mb k}
\left(
\varepsilon_{c,\mb k+\mb Q}
-\varepsilon_{v,\mb k}
\right).
\end{equation}
The interaction part is the direct electron-hole attraction plus exchange,

\begin{equation}
H_{\mb Q}=H^{(1)}_{\mb Q}+H^{\rm dir}_{\mb Q}+H^{\rm ex}_{\mb Q}.
\end{equation}
To clarify the interaction Hamiltonian, the Coulomb interaction in the moir\'e band basis is

\begin{equation}
\hat H^{(2)}
=
\sum_{\mb k_1\mb k_2\mb q}
\sum_{\mb g}
\sum_{\alpha_1\alpha_2\alpha_3\alpha_4}
V_{\alpha_1\alpha_2\alpha_3\alpha_4}
(\mb k_1,\mb k_2,\mb q+\mb g)\,
e^\dagger_{\alpha_1,\mb k_1+\mb q}
e^\dagger_{\alpha_2,\mb k_2-\mb q}
e_{\alpha_3,\mb k_2}
e_{\alpha_4,\mb k_1}.
\end{equation}
the Coulomb matrix is written in terms of form factors as

\begin{equation}
V_{\alpha_1\alpha_2\alpha_3\alpha_4}
(\mb k_1,\mb k_2,\mb q+\mb g)
=
\frac{1}{2\Omega_{\rm tot}}
\Lambda_{\alpha_1\alpha_4}(\mb k_1,\mb q,\mb g)
V(\mb q+\mb g)
\Lambda_{\alpha_2\alpha_3}(\mb k_2,-\mb q,-\mb g),
\end{equation}
or equivalently, using Hermiticity of the form factor,

\begin{equation}
V_{\alpha_1\alpha_2\alpha_3\alpha_4}
(\mb k_1,\mb k_2,\mb q+\mb g)
=
\frac{1}{2\Omega_{\rm tot}}
\Lambda_{\alpha_1\alpha_4}(\mb k_1,\mb q,\mb g)
V(\mb q+\mb g)
\Lambda^*_{\alpha_3\alpha_2}(\mb k_2-\mb q,\mb q,\mb g).
\end{equation}
The form factor describes the scattering between two Bloch eigenstates

\begin{equation}
\Lambda_{\alpha\beta}(\mb p,\mb q,\mb g)
=
\sum_{\mu,\mb G'}
U^*_{\alpha,\mu,\mb G'+\mb g}(\mb p+\mb q)
U_{\beta,\mu,\mb G'}(\mb p),
\end{equation}
where $\mu=A,B$ is the electronic orbital index. 
The form factors satisfy the translation properties,

\begin{equation}
\Lambda_{\alpha\beta}(\mb p+\mb g'',\mb q,\mb g)
=
\Lambda_{\alpha\beta}(\mb p,\mb q,\mb g), \qquad
\Lambda_{\alpha\beta}(\mb p,\mb q+\mb g'',\mb g)
=
\Lambda_{\alpha\beta}(\mb p,\mb q,\mb g+\mb g'').
\end{equation}
for any moir\'e reciprocal lattice vector $\mb g''$, and Hermiticity

\begin{equation}
\Lambda_{\alpha\beta}(\mb p,\mb q,\mb g)
=
\Lambda^*_{\beta\alpha}(\mb p+\mb q,-\mb q,-\mb g).
\end{equation}
For numerical implementation, the final momentum must be folded back into the BZ. Define $[\mb p+\mb q]=\mb p+\mb q-\mb g_0$, the folded form used in numerical calculation is then

\begin{equation}
\Lambda_{\alpha\beta}(\mb p,\mb q,\mb g)
=
\sum_{\mu,\mb G'}
U^*_{\alpha,\mu,\mb G'+\mb g+\mb g_0}([\mb p+\mb q])
U_{\beta,\mu,\mb G'}(\mb p).
\end{equation}
Combining these expressions with the particle-hole BSE basis gives, with $\mb q=\mb k'-\mb k$,

\begin{equation}
H^{\rm dir}_{\mb Q;c'v',cv}(\mb k',\mb k)
=-\frac{1}{\Omega_{\rm tot}}
\sum_{\mb g}
\Lambda_{c'c}(\mb k+\mb Q,\mb q,\mb g)
V_C(\mb q+\mb g)
\left[
\Lambda_{v'v}(\mb k,\mb q,\mb g)
\right]^* ,
\end{equation}
and

\begin{equation}
H^{\rm ex}_{\mb Q;c'v',cv}(\mb k',\mb k)
=\frac{1}{\Omega_{\rm tot}}
\sum_{\mb g}
\Lambda_{c'v'}(\mb k',\mb Q,\mb g)
V_C^{\rm bare}(\mb Q+\mb g)
\left[
\Lambda_{cv}(\mb k,\mb Q,\mb g)
\right]^* .
\end{equation}
The interaction used in the direct term is a double-gated Coulomb form $V_C(\mb q)=(e^2/2\epsilon\epsilon_0)\tanh(|\mb q|d_g)/|\mb q|$, with $\epsilon_0$ being the vacuum permittivity, $d_g=30$nm being the gate distance. 
In exchange term, the interaction is bare Coulomb form $V_C^{\rm bare}$ with $d_g \rightarrow \infty$ and $\epsilon=1$. According to our numerical tests, the results from using the Keldysh form do not alter our conclusion in the main text.

The numerical calculations for the BSE dispersions presented in the main text use a $54 \times 54$ $\mb{k}$-mesh. The single-particle plane-wave basis includes six shells of reciprocal lattice vectors, while the Coulomb-scattering cutoff includes five shells. The phase diagrams of the exciton Chern number and quantum geometry are calculated using a $36 \times 36$ $\mb{k}$-mesh. The band ordering, direct gaps, and Chern numbers remain unchanged upon increasing the momentum-mesh density and reciprocal-space cutoffs.

\subsubsection{Effect of Interaction Strength}
The dielectric constant $\epsilon$ provides a direct control of the electron--hole interaction strength. 
For the screened Coulomb interaction introduced above, the attractive direct term scales as $H^{\rm dir}\propto -1/\epsilon$, while the single-particle Hamiltonian is unchanged. 
Increasing $\epsilon$ therefore weakens the electron--hole attraction and reduces the exciton binding energy, shifting the exciton spectrum toward the single-particle electron--hole continuum. 
At the same time, the change is not simply a rigid energy shift: the relative weight of the single-particle dispersion and the direct interaction in the BSE is modified, so the exciton envelope function $\phi_{n\mb Q}(\mb k)$ also evolves with $\epsilon$. 
In real space, weaker binding generally corresponds to a less localized relative electron--hole wave function, which can modify the effective hopping, bandwidth, and quantum geometry of the resulting CT-exciton bands.

\begin{figure}
\includegraphics[width=1.0\textwidth]{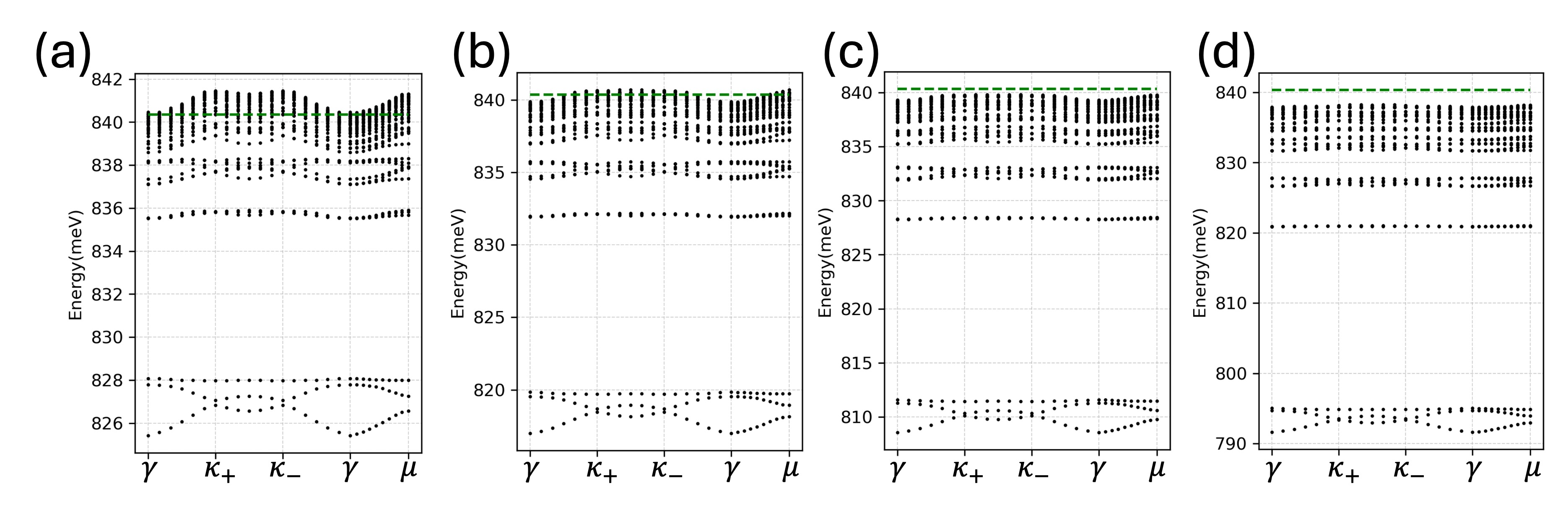}%
\caption{Exciton dispersion showing the lowest 10 BSE bands at different dielectric constants. (a) $\epsilon=30$, (b)$\epsilon=20$, (c)$\epsilon=15$, (d)$\epsilon=10$.
Other parameters are the same as those in Fig. 2(b) in the main text.
The green dashed line marks the single-particle gap. 
\label{figS_dielectric}}
\end{figure}

Figure~\ref{figS_dielectric} compares the BSE dispersions for different dielectric constants. 
Within the parameter range considered here, the low-energy CT-exciton manifold remains well separated from higher excitations and retains its characteristic band structure, while its binding energy and dispersion vary continuously with the interaction strength.
This explains why varying $\epsilon$ in Fig.~3(d) modifies the momentum-space distributions of the exciton Berry curvature and quantum metric without changing the underlying single-particle trapping potential.

\subsubsection{Exchange-Induced Linear Dispersion}
In the main text, BSE spectrum deviates from the tight binding dispersion in the presence of the exchange interaction, which produces the nonanalytic/linear behavior near $\gamma$. 
To see the origin, keep the long-range $\mb g=0$ exchange channel and define the interband overlap

\begin{equation}
M_{cv}(\mb k,\mb Q)
=
\Lambda_{cv}(\mb k,\mb Q,\mb 0)
=
\langle u_{c,\mb k+\mb Q}|u_{v,\mb k}\rangle .
\end{equation}
Orthogonality gives $M_{cv}(\mb k,\mb 0)=0$. A small-$\mb Q$ expansion gives

\begin{equation}
M_{cv}(\mb k,\mb Q)
=
Q_a\langle\partial_{k_a}u_{c,\mb k}|u_{v,\mb k}\rangle
+O(Q^2)
=
\frac{Q_a\langle u_{c,\mb k}|\partial_{k_a}H_0(\mb k)|u_{v,\mb k}\rangle}
{\varepsilon_{c,\mb k}-\varepsilon_{v,\mb k}}
+O(Q^2),
\end{equation}
which is proportional to the optical dipole or velocity matrix element. The exchange kernel contains

\begin{equation}
H^{\rm ex}\sim
V(\mb Q)\,
M_{c'v'}(\mb k',\mb Q)
M^*_{cv}(\mb k,\mb Q).
\end{equation}
For a two-dimensional long-range Coulomb interaction $V_C(\mb Q)\propto1/|\mb Q|$, the two dipole factors contribute $Q^2$, so the bright-sector exchange shift scales as $|\mb Q|$. The short-range and $\mb g\neq0$ exchange components are analytic in $\mb Q$ and give ordinary quadratic corrections.

\section{Strong-Modulation Limit and CT Tight-Binding Model\label{section:strong-modulation}}

\subsection{Exciton wave function in the strong-modulation limit}
In the strong-modulation limit, the lowest conduction and valence bands become flat and are described by maximally localized Wannier orbitals

\begin{equation}
w_{c,\mb R}(\mb r)
=w_c(\mb r-\mb R-\bs{\tau}_c),
\qquad
w_{v,\mb R}(\mb r)
=w_v(\mb r-\mb R-\bs{\tau}_v),
\end{equation}
where $\bs{\tau}_c$ and $\bs{\tau}_v$ are the intracell positions of the conduction and valence Wannier centers. The Bloch wave is

\begin{equation}
\psi_{n,\mb k}(\mb r)
=\frac{1}{\sqrt N}
\sum_{\mb R}
e^{i\mb k\cdot(\mb R+\bs{\tau}_n)}
w_n(\mb r-\mb R-\bs{\tau}_n).
\end{equation}
Because the Wannier orbitals are localized, the direct form factor becomes an on-site form factor times the phase from the Wannier center:

\begin{equation}
\Lambda_n(\mb p,\mb q + \mb g)
=
e^{i\mb g\cdot\bs{\tau}_n}
F_n(\mb q + \mb g),
\qquad
F_n(\mb s)=\int d\mb r\,|w_n(\mb r)|^2e^{i\mb s\cdot\mb r}.
\end{equation}
Here $\mb q=\mb k'-\mb k$ is the BZ momentum transfer. For a single conduction-valence pair, the direct binding kernel is then independent of $\mb Q$ and depends only on $\mb q$:

\begin{equation}
K_{cv}(\mb q)
=
\frac{1}{\Omega_{\rm tot}}
\sum_{\mb g}
F_c(\mb q+\mb g)
V(\mb q+\mb g)
F_v^*(\mb q+\mb g)
e^{i\mb g\cdot(\bs{\tau}_c-\bs{\tau}_v)}.
\end{equation}
Here $K_{cv}$ is the positive binding kernel obtained from the attractive direct term. With the band dispersion flat, the BSE reduces to the convolution equation

\begin{equation}
\sum_{\mb k}
K_{cv}(\mb k'-\mb k)
\phi_n(\mb k)
=E_b\,\phi_n(\mb k'),
\qquad
E_b=\Delta_g-E^{\rm ex}_n.
\end{equation}
Fourier transforming to the relative CT coordinate $\bs{\delta}$,

\begin{equation}
\Phi_n(\bs{\delta})
=\frac{1}{\sqrt N}\sum_{\mb k}
e^{i\mb k\cdot\bs{\delta}}\phi_n(\mb k),
\qquad
K_{cv}(\bs{\delta})
=\sum_{\mb q}
e^{i\mb q\cdot\bs{\delta}}K_{cv}(\mb q),
\end{equation}
gives the real-space Wannier equation

\begin{equation}
K_{cv}(\bs{\delta})\Phi_n(\bs{\delta})
=E_b\Phi_n(\bs{\delta}).
\end{equation}
In the infinitely deep and strongly localized limit, the lowest solution is localized at one of the optimal electron-hole separations $\bs{\delta}_j$:

\begin{equation}
\Phi_j(\bs{\delta})=\delta_{\bs{\delta},\bs{\delta}_j},
\qquad
\phi_j(\mb k)=\frac{1}{\sqrt N}e^{-i\mb k\cdot\bs{\delta}_j}.
\end{equation}
The corresponding localized CT exciton is written in the main-text notation as

\begin{equation}
X_{j,\mb r_i}(\mb r_e,\mb r_h)
=
w_c(\mb r_e-\mb r_i-\bs{\delta}_j)
w_v^*(\mb r_h-\mb r_i),
\end{equation}
where $\mb r_i$ labels the common unit-cell coordinate of the CT-exciton and $\bs{\delta}_j$ labels the relative electron-hole displacement. The associated Bloch operator has the schematic form

\begin{equation}
X^\dagger_{\mb Q,j}
=
\frac{1}{\sqrt N}
\sum_i
e^{i\mb Q\cdot\mb r_i}
X^\dagger_{j,\mb r_i}.
\end{equation}
The three shortest displacements $\bs{\delta}_j$ connect the valence and conduction sublattice positions in the honeycomb configuration.
Under $\mc{PI}$ symmetry, the bond midpoint transforms as the invariant point and provides the symmetry-selected Kagome embeddings, realizing the configuration schematically displayed in Fig.~1(b) of the main text.

\subsection{CT exciton tight-binding model}
Turning on electron and hole hopping perturbatively and restricting to nearest-neighbor processes gives the effective Kagome tight-binding model quoted in the manuscript:

\begin{equation}
H_{\rm eff}
=
\sum_i\sum_{j=1}^3
\left[
t_c X^\dagger_{j+1,\mb r_i}X_{j,\mb r_i}
+
t_v X^\dagger_{j,\mb r_i}X_{j+1,\mb r_i-\mb d_{j+2}}
+{\rm h.c.}
\right],
\end{equation}
with cyclic convention $j+3\equiv j \,\, (\mathrm{mod}\,\,3)$ and
$\mb d_{j+2}=\bs{\delta}_{j+1}-\bs{\delta}_j$.
The effective amplitudes $t_c$ and $t_v$ are generated by conduction-electron and valence-electron hopping, respectively. 
When time-reversal symmetry is broken by using the Dirac kinetic term in BSE, these hopping amplitudes can acquire imaginary parts. When the scalar-potential phase deviates from $\varphi=\pm\pi/2$, the resulting trapping asymmetry produces a real hopping imbalance and breaks $\mc{PI}$.

Let the complex nearest-neighbor hopping amplitudes be

\begin{equation}
\tilde t_c=t_c+i\eta_c,\qquad
\tilde t_v=t_v+i\eta_v,
\qquad
t_c,t_v,\eta_c,\eta_v\in\mathbb R .
\end{equation}
The Bloch Hamiltonian in the three-orbital Kagome basis can be written as

\begin{equation}
H(\mb k)=
\begin{pmatrix}
0&f_3^*(\mb k)&f_2(\mb k)\\
f_3(\mb k)&0&f_1^*(\mb k)\\
f_2^*(\mb k)&f_1(\mb k)&0
\end{pmatrix},
\end{equation}
with

\begin{equation}
f_j(\mb k)=\tilde t_c+\tilde t^*_v e^{i\mb k\cdot\mb d_j},
\qquad j=1,2,3,
\end{equation}
and

\begin{equation}
\mb d_1=a(0,-1),\qquad
\mb d_2=a\left(\frac{\sqrt3}{2},\frac12\right),\qquad
\mb d_3=a\left(-\frac{\sqrt3}{2},\frac12\right).
\end{equation}
It is useful to define

\begin{equation}
t_0=\frac{t_c+t_v}{2},\qquad
\delta t=\frac{t_c-t_v}{2},\qquad
\eta_\Sigma=\frac{\eta_c+\eta_v}{2},\qquad
\delta\eta=\frac{\eta_c-\eta_v}{2}.
\end{equation}
The $C_3$-adapted orbital basis is

\begin{equation}
|s\rangle=\frac{1}{\sqrt3}(1,1,1)^T,\qquad
|+\rangle=\frac{1}{\sqrt3}(1,\omega,\omega^2)^T,\qquad
|-\rangle=\frac{1}{\sqrt3}(1,\omega^2,\omega)^T,
\end{equation}
with $\omega=e^{i2\pi/3}$. In the two-band projections below, $\sigma_0$ is the $2\times2$ identity and $\sigma_{x,y,z}$ are Pauli matrices in the projected two-state subspace.
The unperturbed Kagome limit is $\delta t=\eta_\Sigma=\delta\eta=0$. The relevant avoided crossings are the Dirac points at
$\kappa_{\tau=\pm}=\left(0,\tau\frac{4\pi}{3a}\right)$, and the quadratic touching at $\gamma$.

\subsubsection{Dirac points \texorpdfstring{$\kappa_\tau$}{kappa\_tau}}

Set $\mb k=\bs{\kappa}_\tau+\mb q$, with $qa\ll1$. Projecting the three-band Hamiltonian onto the two-dimensional degenerate subspace gives

\begin{equation}
H^{(\kappa_\tau)}_{\rm eff}
=
\varepsilon_\tau\sigma_0
+v_D(\tau q_x\sigma_x+q_y\sigma_y)
+m_\tau\sigma_z,
\end{equation}
where

\begin{equation}
\varepsilon_\tau=t_0+\tau\sqrt3\,\eta_\Sigma,
\qquad
v_D=\frac{\sqrt3\,a\,t_0}{2},
\end{equation}
and

\begin{equation}
m_\tau
=3\delta t-\tau\sqrt3\,\delta\eta
=\frac{3}{2}(t_c-t_v)
-\tau\frac{\sqrt3}{2}(\eta_c-\eta_v).
\end{equation}
Thus the gap at $\kappa_\tau$ is

\begin{equation}
\Delta_{\kappa_\tau}=2|m_\tau|
=
\left|
3(t_c-t_v)-\tau\sqrt3(\eta_c-\eta_v)
\right|.
\end{equation}
Equivalently, in terms of $\Delta\tilde t=\tilde t_c-\tilde t_v$,

\begin{equation}
\Delta_{\kappa_\tau}
=
\left|
3\,{\rm Re}\,\Delta\tilde t
-\tau\sqrt3\,{\rm Im}\,\Delta\tilde t
\right|.
\end{equation}
The origin of the Dirac-point gap is therefore transparent:

\begin{itemize}
\item $t_c-t_v$ is a real hopping imbalance, corresponding to a breathing-Kagome mass associated with $\mc{PI}$ breaking.
\item $\eta_c-\eta_v$ is the time-reversal-breaking imaginary hopping difference, giving a valley-odd Haldane-type mass.
\item $\eta_c+\eta_v$ shifts the two valleys oppositely, which breaks time reversal symmetry but does not open the Dirac gap at leading order.
\end{itemize}

For the two-band Hamiltonian

\begin{equation}
\mb d_\tau(\mb q)
=
(v_D\tau q_x,\;v_D q_y,\;m_\tau),
\end{equation}
the upper/lower Berry curvature is

\begin{equation}
\Omega_{\tau,\pm}(\mb q)
=
\mp \tau
\frac{m_\tau v_D^2}
{2\left(m_\tau^2+v_D^2q^2\right)^{3/2}}.
\end{equation}
This curvature is sharply concentrated near the avoided Dirac crossing when $|m_\tau|$ is small, consistent with the Berry curvature hot spots near $\kappa_\pm$ in Fig. 2(c) in the main text.

\subsubsection{Quadratic touching at \texorpdfstring{$\gamma$}{gamma}}

Near $\gamma$, write $\mb k=\mb q$, $qa\ll1$, and project onto the $\{|+\rangle,|-\rangle\}$ doublet of the $C_3$-adapted basis. To second order in $\mb q$,

\begin{equation}
H^{(\gamma)}_{\rm eff}
=
-(t_c+t_v)\sigma_0
+\beta q^2\sigma_0
+\beta\left[
(q_x^2-q_y^2)\sigma_x+2q_xq_y\sigma_y
\right]
+m_\gamma\sigma_z,
\end{equation}
where

\begin{equation}
\beta=\frac{a^2t_ct_v}{4(t_c+t_v)},
\qquad
m_\gamma=\sqrt3(\eta_c-\eta_v).
\end{equation}
The $\gamma$-point gap is

\begin{equation}
\Delta_\gamma=2|m_\gamma|
=2\sqrt3|\eta_c-\eta_v|
=\left|2\sqrt3\,{\rm Im}(\tilde t_c-\tilde t_v)\right|.
\end{equation}
Thus $t_c-t_v$ does not open the $\gamma$-point gap at leading order, while the time-reversal-breaking hopping difference $\eta_c-\eta_v$ does. This differs from the Dirac points, where both the real and imaginary hopping differences contribute to the mass.

The $\gamma$-point $d$-vector is

\begin{equation}
\mb d_\gamma(\mb q)
=
\left(
\beta(q_x^2-q_y^2),\;
2\beta q_xq_y,\;
m_\gamma
\right).
\end{equation}
The upper/lower Berry curvature is

\begin{equation}
\Omega_{\gamma,\pm}(\mb q)
=
\mp
\frac{2m_\gamma\beta^2q^2}
{\left(m_\gamma^2+\beta^2q^4\right)^{3/2}}.
\end{equation}
Unlike the Dirac-point curvature, this curvature vanishes at $\mb q=0$ and is distributed on a ring around $\gamma$.

\section{Particle-Hole and Combined Particle-Hole-Inversion Symmetry\label{section:PI}}

This section formulates the particle-hole (PH) symmetry used in the main text, derives the combined PH and inversion symmetry of the continuum model, and gives parallel real-space and momentum-space derivations of its action on charge-transfer (CT) excitons.

\subsection{General particle-hole transformation}

The derivation of PH transformation and PH symmetry follows the literature~\cite{chiu2016classification} with modified convention. We start from the second-quantized convention.
Let

\begin{equation}
\hat\psi=
\begin{pmatrix}
c_1&c_2&\cdots&c_{N_{\rm orb}}
\end{pmatrix}^{T},
\qquad
\hat\psi^\dagger=
\begin{pmatrix}
c_1^\dagger&c_2^\dagger&\cdots&c_{N_{\rm orb}}^\dagger
\end{pmatrix},
\end{equation}
where the index includes all single-particle orbital, position, and internal labels. Let $U_P$ be a unitary matrix in this single-particle basis. We use the convention
\begin{equation}
\mathcal P c_i^\dagger\mathcal P^{-1}
=
\sum_j (U_P^\dagger)_{ij}c_j,
\qquad
\mathcal P c_i\mathcal P^{-1}
=
\sum_j c_j^\dagger(U_P)_{ji}.
\end{equation}
Equivalently,

\begin{equation}
\mathcal P\hat\psi^\dagger\mathcal P^{-1}
=
\left(U_P^\dagger\hat\psi\right)^T,
\qquad
\mathcal P\hat\psi\mathcal P^{-1}
=
\left(\hat\psi^\dagger U_P\right)^T .
\end{equation}
Here $\mathcal P$ denotes the PH operation used in the main text. It is a unitary canonical transformation on Fock space, while its corresponding first-quantized map is antiunitary and can be represented as $U_PK$, with $K$ denoting complex conjugation.
To make the charge-conjugating character explicit, define the fermion number and the charge measured from half filling,
$
\hat N=\sum_i c_i^\dagger c_i,
$
and
$
\hat Q=\hat N-N_{\rm orb}/2.
$
Using the canonical anti-commutation relations and $U_P^\dagger U_P=1$ gives
$
\mathcal P\hat N\mathcal P^{-1}=N_{\rm orb}-\hat N,
$
and
$
\mathcal P\hat Q\mathcal P^{-1}=-\hat Q.
$

The physical electric charge measured from half filling changes sign in the same way. Equivalently, if a hole creation operator is defined relative to the reference sea by $h_i^\dagger\equiv c_i$, then

\begin{equation}
\mathcal P c_i^\dagger\mathcal P^{-1}
=\sum_j(U_P^\dagger)_{ij}h_j^\dagger,
\qquad
\mathcal P h_i^\dagger\mathcal P^{-1}
=\sum_jc_j^\dagger(U_P)_{ji}.
\end{equation}
PH therefore exchanges particle and hole creation operators, with the matrix $U_P$ specifying how their single-particle labels are sewn together.

The action on a quadratic Hamiltonian can be derived by considering

\begin{equation}
\hat H[h]
=
\hat\psi^\dagger h\hat\psi
=
\sum_{ij}c_i^\dagger h_{ij}c_j,
\end{equation}
where $h$ is a Hermitian single-particle matrix. Fermion reordering gives
$
\hat\psi^T A(\hat\psi^\dagger)^T = \mathrm{tr}A-\hat\psi^\dagger A^T\hat\psi
$
for any c-number matrix $A$. It follows that

\begin{equation}
\begin{aligned}
\mathcal P\hat H[h]\mathcal P^{-1}
=
\hat\psi^T U_P^*hU_P^T(\hat\psi^\dagger)^T
=
\operatorname{tr}h
-
\hat\psi^\dagger U_Ph^T U_P^\dagger\hat\psi .
\end{aligned}
\end{equation}
The PH-conjugate one-particle Hamiltonian is therefore
$
h_P=-U_Ph^T U_P^\dagger.
$
The canonical transformation can be defined for any quadratic Hamiltonian, but it is a symmetry only if, after choosing the PH center of energy,

\begin{equation}
U_Ph^T U_P^\dagger=-h.
\end{equation}
The trace term is a c-number and can be absorbed into the many-body energy origin.

\subsection{$\mc{PI}$ symmetry of the continuum model}

\subsubsection{Operator notation}

Below, $a_{\mu,\mathbf r}^\dagger$ creates a microscopic orbital $\mu=A,B$, $e_{\alpha,\mathbf k}^\dagger$ creates a moir\'e Bloch-band state, and $U_{\alpha;\mu\mathbf G}(\mathbf k)$ is a Bloch-wave coefficient defined in Sec.~\ref{section:BSE}.
For the two-orbital continuum model, the combined operation is defined microscopically by

\begin{equation}
(\mathcal{PI})a_{\mu,\mathbf r}^\dagger(\mathcal{PI})^{-1}
=
\sum_\nu(U_{PI})_{\mu\nu}a_{\nu,-\mathbf r},
\qquad
(\mathcal{PI})a_{\mu,\mathbf r}(\mathcal{PI})^{-1}
=
\sum_\nu a_{\nu,-\mathbf r}^\dagger(U^\dagger_{PI})_{\nu\mu},
\end{equation}
In other words, $\mathcal{PI}$ inverts the position, exchanges the two microscopic orbital sectors, and interchanges creation and annihilation operators.
With the Fourier convention

\begin{equation}
a_{\mu,\mathbf r}^\dagger
=
\frac{1}{\sqrt{\Omega}}
\sum_{\mathbf p}e^{i\mathbf p\cdot\mathbf r}
a_{\mu,\mathbf p}^\dagger,
\end{equation}
one obtains

\begin{equation}
(\mathcal{PI})a_{\mu,\mathbf p}^\dagger(\mathcal{PI})^{-1}
=
\sum_\nu(U_{PI})_{\mu\nu}a_{\nu,\mathbf p}.
\end{equation}
Pure PH and spatial inversion each reverse the electronic momentum; their combination therefore preserves the electronic momentum while still exchanging particle and hole sectors.

To obtain the action on the electronic moiré-band basis, write
$
e_{\alpha,\mathbf k}^\dagger
= \sum_{\mu,\mathbf G}
U_{\alpha;\mu\mathbf G}(\mathbf k)
a_{\mu,\mathbf k+\mathbf G}^\dagger
$ 
as in Sec.~\ref{section:BSE}. The corresponding full-band sewing matrix is

\begin{equation}
\mathcal B_{\alpha \beta}(\mathbf k)
=
\sum_{\mu\nu,\mathbf G}
U_{\alpha;\mu\mathbf G}(\mathbf k)
(U_{PI})_{\mu\nu}
U_{\beta;\nu\mathbf G}(\mathbf k).
\end{equation}
It gives

\begin{equation}
(\mathcal{PI})e_{\alpha,\mathbf k}^\dagger(\mathcal{PI})^{-1}
=
\sum_\beta \mathcal B_{\alpha \beta}(\mathbf k)e_{\beta,\mathbf k},
\qquad
(\mathcal{PI})e_{\alpha,\mathbf k}(\mathcal{PI})^{-1}
=
\sum_\beta\mathcal B_{\alpha \beta}^*(\mathbf k)e_{\beta,\mathbf k}^\dagger.
\end{equation}
In strong-modulation limit with the isolated conduction and valence bands, a compatible gauge can be chosen such that $\mathcal B_{cv}\simeq -\mathcal B_{vc}\simeq1$. This results in $\varepsilon_{c,\mb k}=-\varepsilon_{v,\mb k}$ in the whole BZ.

\subsubsection{Constraint on the continuum Hamiltonian}

For the continuum single-particle Hamiltonian defined in Sec.~\ref{section:sp_Hamiltonian}, the orbital part of PH is $U_P=\sigma_x$, while spatial inversion carries the orbital action $U_I=\sigma_z$ in addition to $\mb r \rightarrow -\mb r$. The combined symmetry condition is

\begin{equation}
\sigma_yH_0^T(\mb k,-\mathbf r)\sigma_y=-H_0(\mb k, \mathbf r).
\end{equation}
Defining
$
V_\pm(\mathbf r)
= [V_c(\mathbf r)\pm V_v(\mathbf r)]/2,
$
the symmetry condition becomes

\begin{equation}
V_-(-\mathbf r)=V_-(\mathbf r),
\qquad
V_+(-\mathbf r)=-V_+(\mathbf r),
\end{equation}
For the first-harmonic potentials
$
V_{c/v}(\mathbf r) = 2\Delta_{c/v}\sum_{j=1}^{3}
\cos(\mathbf g_j\cdot\mathbf r-\varphi_{c/v}),
$
this requires
$
\Delta_c e^{i\varphi_c} = -\Delta_v e^{-i\varphi_v}.
$
Taking $\Delta_c,\Delta_v\geq0$ gives

\begin{equation}
\Delta_c=\Delta_v,
\qquad
\varphi_c+\varphi_v=\pi
\pmod{2\pi}.
\end{equation}
The simplified main-text model has $V_c=V_v=V$. Its combined symmetry condition reduces to the odd condition
$
V(-\mathbf r)=-V(\mathbf r).
$
For
$
V(\mathbf r)
= 2\Delta\sum_{j=1}^{3}
\cos(\mathbf g_j\cdot\mathbf r-\varphi),
$
the potential is odd at
$
\varphi=\pm\frac{\pi}{2}\pmod{2\pi},
$
so the main text chooses $\varphi=\pi/2$. This combined PH-inversion symmetry is distinct from ordinary inversion: a particle-number-preserving inversion would require $V_+(-\mathbf r)=V_+(\mathbf r)$ rather than $V_+(-\mathbf r)=-V_+(\mathbf r)$.

\subsection{Action of $\mathcal{PI}$ on the CT-exciton basis}

\subsubsection{Localized real-space excitons in the strong-modulation limit}

Let $\mathbf r_i$ denote the common unit-cell coordinate of an electron-hole pair used in the main text, which is also the valence-orbital center, and let $\boldsymbol\delta_j$ point from the hole endpoint to the electron endpoint. In the strong-modulation limit, let $c_{c,\mathbf r_i+\boldsymbol\delta_j}^\dagger$ create the conduction Wannier orbital centered at $\mathbf r_i+\boldsymbol\delta_j$, and let $c_{v,\mathbf r_i}$ remove an electron from the valence Wannier orbital centered at $\mathbf r_i$. The localized CT-exciton operator is
\begin{equation}
X_{j,\mathbf r_i}^\dagger = c_{c,\mathbf r_i+\boldsymbol\delta_j}^\dagger c_{v,\mathbf r_i}.
\end{equation}
Its real-space wave function is the form used in the main text,
\begin{equation}
X_{j,\mathbf r_i}(\mathbf r_e,\mathbf r_h) = w_c(\mathbf r_e-\mathbf r_i-\boldsymbol\delta_j) w_v^*(\mathbf r_h-\mathbf r_i).
\end{equation}

At the exact $\mathcal{PI}$-symmetric point, the conduction and valence Wannier orbitals are PH partners. Their phases can be chosen so that

\begin{equation}
(\mathcal{PI})
c_{c,\mathbf r_i+\boldsymbol\delta_j}^\dagger
(\mathcal{PI})^{-1}
=
c_{v,-\mathbf r_i-\boldsymbol\delta_j},
\qquad
(\mathcal{PI})c_{v,\mathbf r_i}(\mathcal{PI})^{-1}
=
-c_{c,-\mathbf r_i}^\dagger .
\end{equation}
The first relation inverts the conduction-orbital center and converts its electron creation operator into valence-electron annihilation; the second relation acts conversely on the valence orbital. Fermion reordering then gives

\begin{equation}
(\mathcal{PI})X_{j,\mathbf r_i}^\dagger(\mathcal{PI})^{-1}
=
-c_{v,-\mathbf r_i-\boldsymbol\delta_j} c_{c,-\mathbf r_i}^\dagger
=
c_{c,-\mathbf r_i}^\dagger c_{v,-\mathbf r_i-\boldsymbol\delta_j}
=
X_{j,-\mathbf r_i-\boldsymbol\delta_j}^\dagger.
\end{equation}
The minus sign from transformation cancels with that from fermion-reordering.
The three shortest displacements $\boldsymbol\delta_j$, $j=1,2,3$, are permuted by $\mathcal C_{3z}$ but are individually preserved by $\mathcal{PI}$ in the endpoint-exchange sense above. They therefore define the three bond-centered CT orbitals of the effective Kagome manifold.

\subsubsection{Momentum-space derivation from the BSE exciton state}

The momentum-space transformation can be derived directly from the BSE exciton state. Using the convention defined in Sec.~\ref{section:BSE},

\begin{equation}
|X_{n,\mathbf Q}\rangle
=
\sum_{\mathbf k,cv}
\phi_{n\mathbf Q,cv}(\mathbf k)
b_{\mathbf Q,\mathbf k;cv}^\dagger
|{\rm GS}\rangle.
\end{equation}
and using the electronic band-basis action and sewing matrix $\mathcal B(\mathbf k)$ derived above, the pair operator transforms as

\begin{equation}
\begin{aligned}
(\mathcal{PI})b_{\mathbf Q,\mathbf k;cv}^\dagger(\mathcal{PI})^{-1}
&=
\sum_{\alpha\beta}
\mathcal B_{c \alpha}(\mathbf k+\mathbf Q)
\mathcal B_{v\beta}^*(\mathbf k)
e_{\alpha,\mathbf k+\mathbf Q}
e_{\beta,\mathbf k}^\dagger\\
&=
-
\sum_{c' v'}
\mathcal B_{c v'}(\mathbf k+\mathbf Q)
\mathcal B_{v c'}^*(\mathbf k)
e_{c',\mathbf k}^\dagger
e_{v',\mathbf k+\mathbf Q}\\
&=
\sum_{c' v'}
\left[\Pi_{\mathbf Q}(\mathbf k)\right]_{c' v',cv}
b_{-\mathbf Q,\mathbf k+\mathbf Q;c' v'}^\dagger .
\end{aligned}
\end{equation}
Exact $\mathcal{PI}$ symmetry maps the source conduction and valence subspaces into orthogonal target valence and conduction subspaces, respectively, so the fermion reordering above produces no contraction term. The pair-space sewing matrix is

\begin{equation}
\left[\Pi_{\mathbf Q}(\mathbf k)\right]_{c' v',cv}
=
-\mathcal B_{c v'}(\mathbf k+\mathbf Q)
\mathcal B_{v c'}^*(\mathbf k).
\end{equation}
The indices $c'$ and $v'$ label the target conduction and valence subspaces after PH exchanges the source conduction and valence sectors. Assume that the reference ground state is symmetry covariant,
$
(\mathcal{PI})|{\rm GS}\rangle = e^{i\chi_\Pi}|{\rm GS}\rangle ,
$
acting on the momentum-space exciton state and setting
$\bar{\mathbf k}=\mathbf k+\mathbf Q$ gives

\begin{equation}
(\mathcal{PI})|X_{n,\mathbf Q}\rangle
=
e^{i\chi_\Pi}
\sum_{\bar{\mathbf k},c' v'}
\phi^\Pi_{n,-\mathbf Q,c' v'}
(\bar{\mathbf k})
b_{-\mathbf Q,\bar{\mathbf k};c' v'}^\dagger
|{\rm GS}\rangle,
\end{equation}
where

\begin{equation}
\phi^\Pi_{n,-\mathbf Q,c' v'}
(\bar{\mathbf k})
=
\sum_{cv}
\left[
\Pi_{\mathbf Q}(\bar{\mathbf k}-\mathbf Q)
\right]_{c' v',cv}
\phi_{n\mathbf Q,cv}
(\bar{\mathbf k}-\mathbf Q).
\end{equation}
$\mathcal{PI}$ is implemented here as a unitary canonical transformation on Fock space, because it does not conjugate $\phi$.

In strong-modulation limit with the isolated conduction and valence bands, a compatible gauge can be chosen such that $\mathcal B_{cv}\simeq -\mathcal B_{vc}\simeq1$. The pair and envelope transformations then reduce to

\begin{equation}
(\mathcal{PI})b_{\mathbf Q,\mathbf k}^\dagger(\mathcal{PI})^{-1}
\simeq
b_{-\mathbf Q,\mathbf k+\mathbf Q}^\dagger,
\end{equation}
and

\begin{equation}
\phi^\Pi_{n,-\mathbf Q}
(\bar{\mathbf k})
\simeq
\phi_{n\mathbf Q}
(\bar{\mathbf k}-\mathbf Q).
\end{equation}

\subsubsection{Induced constraint on the BSE}

Write the BSE Hamiltonian as

\begin{equation}
\hat H_{\rm BSE}
=
\sum_{\mathbf Q}
\sum_{c'v'\mathbf k'}
\sum_{cv\mathbf k}
b_{\mathbf Q,\mathbf k';c'v'}^\dagger
\left[H_{\mathbf Q}(\mathbf k',\mathbf k)\right]_{c'v';cv}
b_{\mathbf Q,\mathbf k;cv}.
\end{equation}
The envelope appearing in the momentum-space exciton state satisfies

\begin{equation}
\sum_{cv\mathbf k}
\left[H_{\mathbf Q}(\mathbf k',\mathbf k)\right]_{c'v';cv}
\phi_{n\mathbf Q,cv}(\mathbf k)
=
E_n(\mathbf Q)
\phi_{n\mathbf Q,c'v'}(\mathbf k').
\end{equation}
Under the same ground-state covariance assumed above, and provided that the retained electron-hole subspace is closed under $\mathcal{PI}$ and that the direct and exchange kernels respect the same microscopic symmetry, the induced BSE covariance is

\begin{equation}
\begin{aligned}
\left[
H_{-\mathbf Q}(\mathbf k'+\mathbf Q,\mathbf k+\mathbf Q)
\right]_{c'v';cv}
=
\sum_{c_1 v_1 c_2 v_2}
\left[\Pi_{\mathbf Q}(\mathbf k')\right]_{c'v';c_1 v_1}
\left[H_{\mathbf Q}(\mathbf k',\mathbf k)\right]_{c_1 v_1;c_2 v_2}
\left[\Pi_{\mathbf Q}^\dagger(\mathbf k)\right]_{c_2 v_2, cv}.
\end{aligned}
\end{equation}
This equation is the precise relation between the $\mathbf Q$ and $-\mathbf Q$ BSE blocks in the numerical momentum convention.
Therefore, at generic $\mathbf Q$, the two states are distinct symmetry partners at $\mathbf Q$ and $-\mathbf Q$. Symmetry does not mean that each state is invariant.
It enforces equal spectra at the two momenta, possibly with a permutation among exciton-band indices: $E_n(\mathbf Q)=E_{n'}(-\mathbf Q)$.

Finally, in a multiband calculation or a general Bloch gauge, the matrices $\mathcal B(\mathbf k)$ and $\Pi_{\mathbf Q}(\mathbf k)$ are essential. A test based only on the momentum relabeling $\mathbf k\mapsto\mathbf k+\mathbf Q$ is not gauge covariant and need not vanish even when the underlying continuum model has exact $\mathcal{PI}$ symmetry.

\section{Exciton Berry Connection, Curvature, and Wilson-Line Evaluation}

In this section we derive the exciton quantum geometry, in both analytic and numerical ways. The real-space exciton wave function can be written as

\begin{equation}
X_{n,\mb Q}(\mb r_e,\mb r_h)
=e^{i\mb Q\cdot\mb R}
U_{n,\mb Q}(\mb r_e,\mb r_h),
\end{equation}
where $\mb R$ is the COM coordinate used to define the exciton Bloch phase. The main text defines the exciton quantum geometric tensor as

\begin{equation}
[g_{ab}(\mb Q)]_n
=
\langle \partial_a U_{n,\mb Q}|
\left(1-|U_{n,\mb Q}\rangle\langle U_{n,\mb Q}|\right)
|\partial_b U_{n,\mb Q}\rangle,
\qquad
\partial_a\equiv\partial_{Q_a}.
\end{equation}
With the Berry connection convention $A^a_n(\mb Q)=i\langle U_{n,\mb Q}|\partial_a U_{n,\mb Q}\rangle$, the Berry curvature is

\begin{equation}
\Omega^z_n(\mb Q)
=\partial_{Q_x}A^y_n(\mb Q)-\partial_{Q_y}A^x_n(\mb Q)
=-2\,{\rm Im}\,[g_{xy}(\mb Q)]_n,
\end{equation}
and the quantum metric is the real part,
$G^n_{ab}(\mb Q)={\rm Re}\,[g_{ab}(\mb Q)]_n$.

The periodic part $U_{n,\mb Q}$ contains the envelope function and the periodic Bloch functions $u_{c,\mb k_e}$, $u_{v,\mb k_h}$. Define the non-Abelian band Berry connections

\begin{equation}
\mc A^a_{\alpha'\alpha}(\mb p)
=i\langle u_{\alpha',\mb p}|\partial_{p_a}u_{\alpha,\mb p}\rangle .
\end{equation}
Then the exciton Berry connection is

\begin{equation}
\begin{aligned}
A^a_n(\mb Q)
=&
\sum_{cv\mb k}
\phi^*_{n\mb Q,cv}(\mb k)
i\partial_{Q_a}\phi_{n\mb Q,cv}(\mb k)\\
&+\gamma_e
\sum_{c'cv\mb k}
\phi^*_{n\mb Q,c'v}(\mb k)
\phi_{n\mb Q,cv}(\mb k)
\mc A^a_{c'c}(\mb k+\gamma_e\mb Q)\\
&+\gamma_h
\sum_{cv'v\mb k}
\phi^*_{n\mb Q,cv'}(\mb k)
\phi_{n\mb Q,cv}(\mb k)
\mc A^{a*}_{v'v}(\mb k-\gamma_h\mb Q).
\end{aligned}
\end{equation}
where mass weight is defined by $\gamma_{e/h}=m_{e/h}/(m_e+m_h)$.  For a single isolated conduction-valence pair this reduces to

\begin{equation}
\mb A_n(\mb Q)
=
\sum_{\mb k}
\phi^*_{n\mb Q}(\mb k)
i\nabla_{\mb Q}\phi_{n\mb Q}(\mb k)
+
\sum_{\mb k}|\phi_{n\mb Q}(\mb k)|^2
\left[
\gamma_e\mb A_c(\mb k+\gamma_e\mb Q)
+\gamma_h\mb A_v(\mb k-\gamma_h\mb Q)
\right].
\end{equation}
The curvature follows by taking $\nabla_{\mb Q}\times\mb A_n$. Equivalently, for nondegenerate bands it can be computed directly from the eigenstate derivative formula

\begin{equation}
\begin{aligned}
\Omega^z_n(\mb Q)
=&\sum_{\mb k} i[\partial_{Q_x}\phi^*_{n\mb Q}(\mb k)\partial_{Q_y}\phi_{n\mb Q}(\mb k)-\partial_{Q_y}\phi^*_{n\mb Q}(\mb k)\partial_{Q_x}\phi_{n\mb Q}(\mb k)] \\
&+|\phi_{n \mb Q}(\mb k)|^2[\gamma_e^2 \Omega^z_c(\mb k + \gamma_e \mb Q) - \gamma_h^2 \Omega^z_v(\mb k - \gamma_h \mb Q)] \\
&+ [\partial_{Q_x}|\phi_{n \mb Q}(\mb k)|^2\gamma_e A^y_c(\mb k + \gamma_e \mb Q) + \partial_{Q_x}|\phi_{n \mb Q}(\mb k)|^2\gamma_h A^y_v(\mb k - \gamma_h \mb Q) - (x \leftrightarrow y)].    
\end{aligned}
\end{equation}

For numerical evaluation, it is inconvenient to deal with the Berry connections and the derivative of the envelope function, so we define the gauge-stable exciton Wilson line

\begin{equation}
W^{\rm ex}_{n,\mb Q}(\delta\mb Q)
=\langle U_{n,\mb Q+\delta\mb Q}|U_{n,\mb Q}\rangle .
\end{equation}
Define the electronic Wilson matrix by its band indices alone:
$
\left[W(\mb p,\delta\mb p)\right]_{\alpha'\alpha}
=
\langle u_{\alpha',\mb p+\delta\mb p}|u_{\alpha,\mb p}\rangle
$.
Then the exciton Wilson line is

\begin{equation}
W^{\rm ex}_{n,\mb Q}(\delta\mb Q)
=
\sum_{\bar c\bar v cv\mb k}
\phi^*_{n,\mb Q+\delta\mb Q,\bar c\bar v}(\mb k)
\left[W(\mb k+\gamma_e\mb Q,\gamma_e\delta\mb Q)\right]_{\bar c c}
\left[W(\mb k-\gamma_h\mb Q,-\gamma_h\delta\mb Q)\right]^*_{\bar v v}
\phi_{n,\mb Q,cv}(\mb k).
\end{equation}
The numerical BSE basis used in Sec.~\ref{section:BSE} corresponds to the momentum-label convention $(\gamma_e,\gamma_h)=(1,0)$. This is not a physical mass assumption; it is related to any other partition by a change of relative momentum. In that convention,

\begin{equation}
W^{\rm ex}_{n,\mb Q}(\delta\mb Q)
=
\sum_{\bar c c v\mb k}
\phi^*_{n,\mb Q+\delta\mb Q,\bar c v}(\mb k)
\left[W(\mb k+\mb Q,\delta\mb Q)\right]_{\bar c c}
\phi_{n,\mb Q,cv}(\mb k).
\end{equation}
Let $\tilde W^{\rm ex}=W^{\rm ex}/|W^{\rm ex}|$. The Berry flux through an elementary plaquette spanned by $\delta Q_x\hat x$ and $\delta Q_y\hat y$ is

\begin{equation}
\Omega^z_n(\mb Q)\,\delta Q_x\delta Q_y
=\operatorname{Im}\log
\left[
\tilde W^{\rm ex}_{n,\mb Q}(\delta Q_x\hat x)
\tilde W^{\rm ex}_{n,\mb Q+\delta Q_x\hat x}(\delta Q_y\hat y)
\left[\tilde W^{\rm ex}_{n,\mb Q+\delta Q_y\hat y}(\delta Q_x\hat x)\right]^{-1}
\left[\tilde W^{\rm ex}_{n,\mb Q}(\delta Q_y\hat y)\right]^{-1}
\right].
\end{equation}
This equation fixes the curvature sign convention by the displayed plaquette orientation. Reversing the Wilson-line order or the coordinate orientation flips the sign of the extracted Berry curvature and Chern number.

To obtain exciton quantum metric, we first expand the neighboring Bloch states,
\begin{equation}
|U_{n,\mathbf Q+\delta\mathbf Q}\rangle
=
|U_{n,\mb Q}\rangle
+
\delta Q_a|\partial_a U_{n,\mb Q}\rangle
+
\frac{\delta Q_a^2}{2}
|\partial_a^2 U_{n,\mb Q}\rangle+\cdots .
\end{equation}
Using the normalization,
\begin{equation}
\partial_a\langle U_{n,\mb Q}|U_{n,\mb Q}\rangle=0,
\qquad
\partial_a^2\langle U_{n,\mb Q}|U_{n,\mb Q}\rangle=0,
\end{equation}
one obtains
\begin{equation}
|W^{\rm ex}_{n,\mb Q}(\delta Q_a\hat a)|^2
=
1-
\left[
\langle\partial_a U_{n,\mb Q}|\partial_a U_{n,\mb Q}\rangle
-
|\langle U_{n,\mb Q}|\partial_a U_{n,\mb Q}\rangle|^2
\right](\delta Q_a)^2
+
O(\delta Q_a^3).    
\end{equation}
The term in brackets is precisely
\begin{equation}
G_{aa}^n(\mathbf Q)
=
\operatorname{Re}
\langle\partial_a U_{n,\mb Q}|
(1-|U_{n,\mb Q}\rangle\langle U_{n,\mb Q}|)
|\partial_a U_{n,\mb Q}\rangle   
\end{equation}
Therefore, the quantum metric can be extracted from the Wilson-line magnitude:
\begin{equation}
G^n_{aa}(\mb Q)(\delta Q_a)^2
=-\log|W^{\rm ex}_{n,\mb Q}(\delta Q_a\hat a)|^2
+O(|\delta\mb Q|^3).
\end{equation}
For a non-orthogonal reciprocal mesh, let $\mb g_1,\mb g_2$ be the primitive reciprocal lattice vectors spanning the BZ and let $\mathsf M=(\mb g_1\ \mb g_2)$ be the $2\times2$ matrix with these vectors as columns. Write $\mb Q=\mathsf M\mb x$, with fractional coordinates $\mb x=(x_1,x_2)$. The Wilson-line finite differences determine the fractional-coordinate metric $\tilde G$ through

\begin{equation}
-\log|W^{\rm ex}_{n,\mb Q}(\delta x_i\,\mb g_i)|^2
=
\tilde G_{ii}(\mb x)(\delta x_i)^2
+O(\delta x_i^3),
\end{equation}
and

\begin{equation}
-\log|W^{\rm ex}_{n,\mb Q}(\delta x(\mb g_1+\mb g_2))|^2
=
\left[
\tilde G_{11}+\tilde G_{22}+2\tilde G_{12}
\right](\delta x)^2
+O(\delta x^3).
\end{equation}
Since $d\mb Q=\mathsf M\,d\mb x$,

\begin{equation}
\tilde G(\mb x)=\mathsf M^T G(\mb Q)\mathsf M,
\qquad
G(\mb Q)=\mathsf M^{-T}\tilde G(\mb x)\mathsf M^{-1}.
\end{equation}
Thus the trace entering the quantum-metric integral is

\begin{equation}
\operatorname{tr}G(\mb Q)
=
\operatorname{tr}
\left[
\mathsf M^{-T}\tilde G(\mb x)\mathsf M^{-1}
\right],
\end{equation}
and

\begin{equation}
\int_{\rm BZ}\frac{d^2Q}{2\pi}\operatorname{tr}G(\mb Q)
=
\frac{|\det\mathsf M|}{2\pi}
\int_{[0,1)^2}d^2x\,
\operatorname{tr}
\left[
\mathsf M^{-T}\tilde G(\mb x)\mathsf M^{-1}
\right].
\end{equation}
This is the numerical procedure underlying the Berry curvature and quantum metric in Fig. 2(c) of the main text.

\end{document}